\documentclass[twocolumn]{aastex63}

\usepackage{pifont}
\usepackage{amsmath}
\usepackage{enumitem}
\usepackage{tablefootnote}
\usepackage{multirow}
\usepackage{color, colortbl}
\definecolor{Gray}{gray}{0.9}

\newcommand{\xmark}{\ding{55}}

\received{\today}

\submitjournal{ApJ}
\received{October 29, 2024}
\accepted{February 6, 2025}
\shorttitle{Scylla IV: BEAST}
\graphicspath{{./}{figures/}}

\begin{document}

\title{Scylla IV: Intrinsic Stellar Properties and Line-of-Sight Dust Extinction Measurements Towards 1.5 Million Stars in the SMC and LMC}

\correspondingauthor{Christina W. Lindberg}
\email{christina.lindberg@live.com}

\author[0000-0003-0588-7360]{Christina W. Lindberg}
\affil{The William H. Miller III Department of Physics \& Astronomy, Bloomberg Center for Physics and Astronomy, Johns Hopkins University, 3400 N. Charles Street, Baltimore, MD 21218, USA}
\affil{Space Telescope Science Institute, 
3700 San Martin Drive, 
Baltimore, MD 21218, USA}

\author[0000-0002-7743-8129]{Claire E. Murray}
\affil{Space Telescope Science Institute, 
3700 San Martin Drive, 
Baltimore, MD 21218, USA}
\affil{The William H. Miller III Department of Physics \& Astronomy, Bloomberg Center for Physics and Astronomy, Johns Hopkins University, 3400 N. Charles Street, Baltimore, MD 21218, USA}

\author[0000-0002-9912-6046]{Petia Yanchulova Merica-Jones}
\affil{Space Telescope Science Institute, 
3700 San Martin Drive, 
Baltimore, MD 21218, USA}


\author[0000-0001-6118-2985]{Caroline Bot}
\affil{Observatoire Astronomique de Strasbourg, Universit\'e de Strasbourg, UMR 7550, 11 rue de l'Universit\'e, F-67000 Strasbourg, France}

\author[0009-0005-0339-015X]{Clare Burhenne}
\affil{Department of Physics and Astronomy, Rutgers the State University of New Jersey, 136 Frelinghuysen Rd., Piscataway, NJ, 08854, USA}

\author[0000-0003-1680-1884]{Yumi Choi}
\affil{NSF National Optical-Infrared Astronomy Research Laboratory, 950 N. Cherry Avenue, Tucson, AZ 85719 USA}

\author[0000-0001-7959-4902]{Christopher J. R. Clark}
\affil{AURA for the European Space Agency, Space Telescope Science Institute, 3700 San Martin Drive, Baltimore, MD 21218, USA}

\author[0000-0002-2970-7435]{Roger E. Cohen}
\affil{Department of Physics and Astronomy, Rutgers the State University of New Jersey, 136 Frelinghuysen Rd., Piscataway, NJ, 08854, USA}

\author[0000-0003-0394-8377]{Karoline M. Gilbert}
\affil{Space Telescope Science Institute, 3700 San Martin Drive, Baltimore, MD 21218, USA}
\affil{The William H. Miller III Department of Physics \& Astronomy, Bloomberg Center for Physics and Astronomy, Johns Hopkins University, 3400 N. Charles Street, Baltimore, MD 21218, USA}

\author[0000-0002-8937-3844]{Steven R. Goldman}
\affil{Space Telescope Science Institute, 3700 San Martin Drive, Baltimore, MD 21218, USA}

\author[0000-0001-5340-6774]{Karl D.\ Gordon}
\affil{Space Telescope Science Institute, 3700 San Martin Drive, Baltimore, MD 21218, USA}

\author[0000-0002-2954-8622]{Alec S.\ Hirschauer}
\affil{Department of Physics \& Engineering Physics, Morgan State University, 1700 East Cold Spring Lane, Baltimore, MD 21251, USA}
\affil{Space Telescope Science Institute, 3700 San Martin Drive, Baltimore, MD 21218, USA}

\author[0000-0001-5538-2614]{Kristen B. W. McQuinn}
\affil{Department of Physics and Astronomy, Rutgers the State University of New Jersey, 136 Frelinghuysen Rd., Piscataway, NJ, 08854, USA}
\affil{Space Telescope Science Institute, 3700 San Martin Drive, 
Baltimore, MD 21218, USA}

\author[0000-0001-6326-7069]{Julia C. Roman-Duval}
\affil{Space Telescope Science Institute, 3700 San Martin Drive, Baltimore, MD 21218, USA}

\author[0000-0002-4378-8534]{Karin M. Sandstrom}
\affil{Department of Astronomy \& Astrophysics, University of California San Diego, 9500 Gilman Drive, La Jolla, CA 92093, USA}

\author[0000-0003-1356-1096]{Elizabeth Tarantino}
\affil{Space Telescope Science Institute, 3700 San Martin Drive, Baltimore, MD 21218, USA}

\author[0000-0002-7502-0597]{Benjamin F. Williams}
\affil{Department of Astronomy, University of Washington, Box 351580, U. W., Seattle, WA 98195-1580, USA}

\begin{abstract}
By analyzing the spectral energy distributions (SEDs) of resolved stars in nearby galaxies, we can constrain their stellar properties and line-of-sight dust extinction. 
From the Scylla survey, we obtain ultraviolet to near-infrared photometry from Wide Field Camera 3 onboard the {\it Hubble Space Telescope} for more than 1.5 million stars in the SMC and LMC. We use the Bayesian Extinction and Stellar Tool (BEAST) to analyze the multi-band SEDs of these sources and characterize their initial masses, ages, metallicities, distances, and line-of-sight extinction properties (e.g.~$A_V$, $R_V$).
We apply quality cuts and perform validation simulations to construct a catalog of over 550,000 stars with high-reliability SED fits, which we use to analyze the stellar content and extinction properties of the SMC and LMC.
We detect stars with masses as low as 0.6 $M_{\odot}$.
BEAST stellar age distributions show a jump in observed stars around 6 Gyrs ago, which agrees with star-formation histories. Extinctions ($A_V$) in both galaxies follow a log-normal distribution.
We compare $A_V$ with ancillary gas and dust tracers like $HI$, $H\alpha$, and far infrared (FIR) dust emission and find positive correlations on a field-by-field basis. We convert observed $A_V$ to predicted dust surface densities using the Draine et. al. (2014) model and find $A_V$-based dust surface densities are a factor of $\sim$2.5 lower than observed FIR-based dust surface densities, a correction factor similar to other studies.

\end{abstract}

\keywords{Magellanic Clouds, Hubble Space Telescope, Stellar Populations, Interstellar Medium}

\section{Introduction} 
\label{sec:intro}

Studying resolved stellar populations in nearby galaxies provides a powerful window into the processes that drive galaxy formation and evolution. By observing individual stars over a range of wavelengths from ultraviolet (UV) to infrared (IR), we can infer their physical properties such as mass, age, metallicity, and the effects of dust extinction. These measurements enable a diverse range of science, including reconstructing galaxy assembly histories from stellar age/metallicity distributions \citep[e.g.,][]{mcquinn2010}; spatio-temporal mapping of star formation patterns \citep[e.g.,][]{lewis2015, williams2017}; direct constraints on stellar feedback processes \citep[e.g.,][]{choi2020}; mapping of dust properties via individual extinction measurements \citep[e.g.,][]{lindberg2024}; and detailed studies of star cluster formation, dynamics, and mass functions by resolving individual members \citep[e.g.,][]{johnson2017}.

Deriving the physical properties of resolved stars has often been done with detailed spectral line information. A high signal-to-noise stellar spectrum can reveal its precise chemical composition, temperature, surface gravity \citep[e.g.,][]{ting2019}, rotational velocity \citep[e.g.,][]{huang2010} and magnetic field \citep[e.g.,][]{leone2017}. However, these observations are expensive to obtain, and the required detailed modeling and analysis can be computationally complex. An alternative approach is to model the broadband spectral energy distribution (SED) of resolved stars \citep[e.g.~][]{conroy2013}. Although this approach cannot constrain the same level of detail on the atmospheric properties of a star as spectral modeling \citep[e.g.,][]{Telford2024}, broad-band SED fitting is consistent with spectral-based fitting \citep{gull2022} and is observationally more efficient, making it ideally suited for analyzing large stellar surveys (e.g.~$10^3\mbox{--}10^6$ sources) of nearby galaxies.

Although numerous SED modeling techniques exist, they generally focus on fitting Galactic sources which suffer minimal crowding compared to resolved star surveys in external galaxies \citep{bailerjones2011, green2014, schonrich2014, ness2015}. Overcoming these limitations is essential to take full advantage of the rich, but complex datasets provided by current and upcoming resolved stellar population surveys targeting nearby galaxies; e.g.~M31: PHAT \citep{dalcanton2012}, PHAST \citep{phast2021hst}, M33: PHATTER \citep{williams2021}, SMC: SMIDGE \citep{ymj2017}, LMC: HTTP \citep{sabbi2016}, METAL \citep{romanduval2019}, LUVIT \citep{gilbert2024}, etc. This motivated the development of the probabilistic Bayesian Extinction and Stellar Tool \citep[BEAST;][]{gordon2016} to robustly measure stellar and dust properties of resolved stellar populations in nearby galaxies while precisely characterizing parameter uncertainties across all observed signal-to-noise regimes.
 

The LMC and SMC (collectively as the MCs) are nearby dwarf galaxies at 50 and 60 kpc, respectively \citep[][]{pietrzynski2019, scowcroft2016}, whose close proximity provide an opportune chance to study resolved stars in external galaxies. 
These dwarf galaxies are likely on their first infall into the Milky Way \citep{besla2007, bekki2009}, but have been interacting with each other for several Gyrs \citep{patel2020}. Based on stellar orbital modeling, it is thought that these galaxies likely suffered a direct collision as recently as 140-160 Myr ago \citep{choi2022, Cullinane2022}, the effects of which are still kinematically, morphologically \citep[e.g.][]{nidever2008, mastropietro2009, tsuge2019} and chemically \citep[e.g.][]{fox2018, romanduval2021} observable today, making them a valuable case study of dwarf-dwarf interactions and enhanced star formation.

The SMC and LMC have experienced relatively less chemical enrichment over their lifetimes compared to the Milky Way, with metallicities of $0.2\ Z_{\odot}$ and $0.5\ Z_{\odot}$, respectively \citep{russell1992}. This makes them excellent laboratories for studying the effects of low metallicity on small-scale physics: stellar evolution, e.g.~mass-loss rates in winds and ionizing radiation from massive stars \citep{crowther2019, telford2023, evans2023, Telford2024, Tarantino2024, hawcroft2024, Ramachandran2024}; AGB dust production \citep{boyer2012, Goldman2022}; and the interstellar medium (ISM), e.g.~dust extinction \citep{gordon2003, gordon2024}, composition and dust-to-gas ratio \citep{Clark2023, gordon2014, romanduval2014, romanduval2017, romanduval2021, romanduval2022a}, molecular gas formation \citep{bolatto2011, jameson2016}.

With new \emph{Hubble Space Telescope} (HST) imaging from the Scylla survey \citep{murray2024b}, we have multi-band photometry for nearly 100 fields across the SMC and LMC, sampling a diverse range of environments, metallicities, gas column densities, and radiation fields. Deep, high-resolution photometry means we can resolve stars down to sub-solar masses \citep{sabbi2013}, giving us unprecedented insight into the stellar content of these fields \citep{cohen2024a, cohen2024b}, especially compared to crowding-limited ground-based surveys like the VISTA survey of the MC system \citep[VMC;][]{cignoni2013} and APOGEE DR17 \citep{Majewski2017, Zasowski2017, Abdurrouf2022}, as well as the flux-limited space-based Gaia survey \citep{gaiacollaboration2018}. By characterizing resolved stars and their line-of-sight extinction in the MCs, we can gain information about the stellar and dust content in these galaxies.

\begin{figure*}
 	\centering
\includegraphics[width=\textwidth]{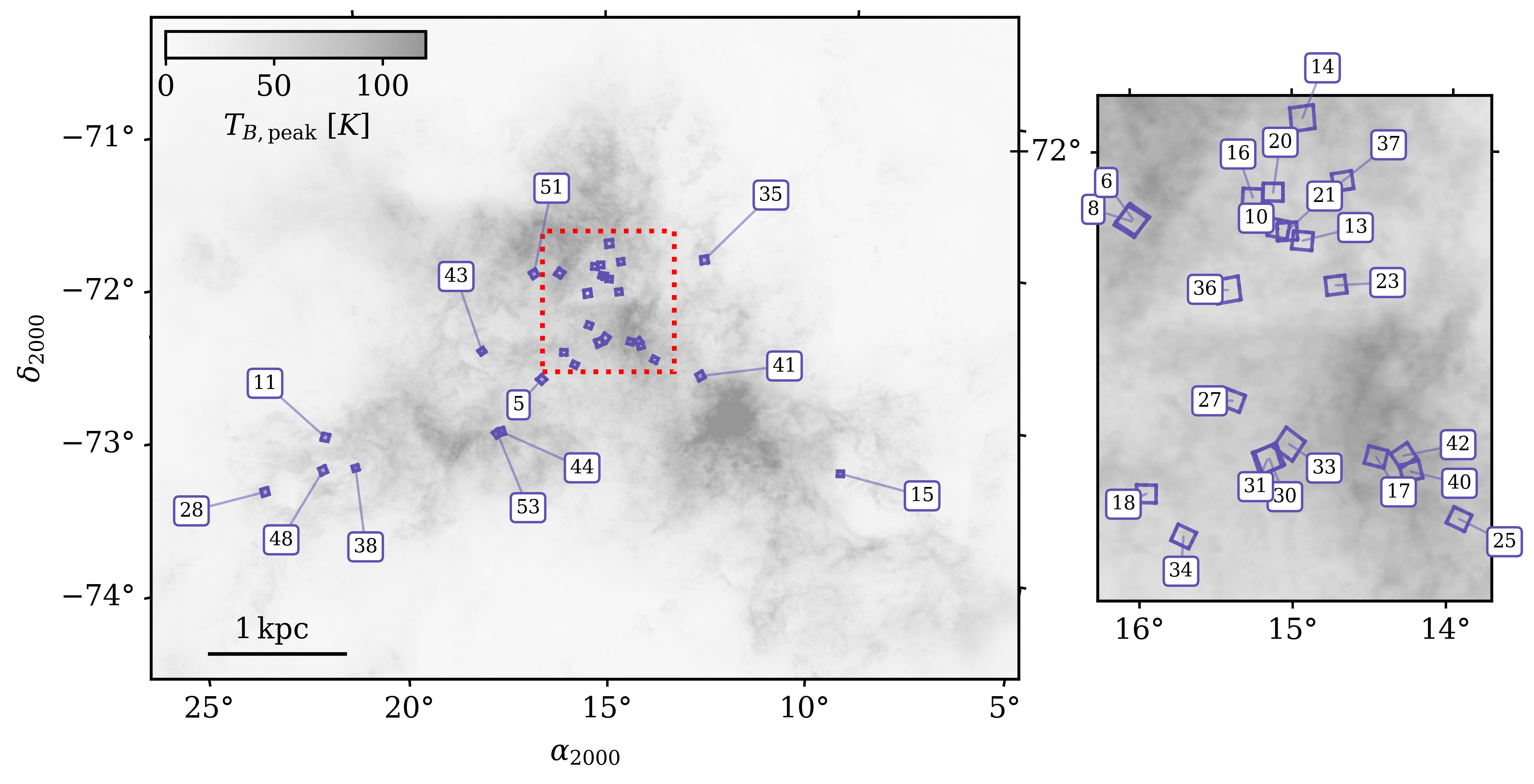}
\includegraphics[width=\textwidth]{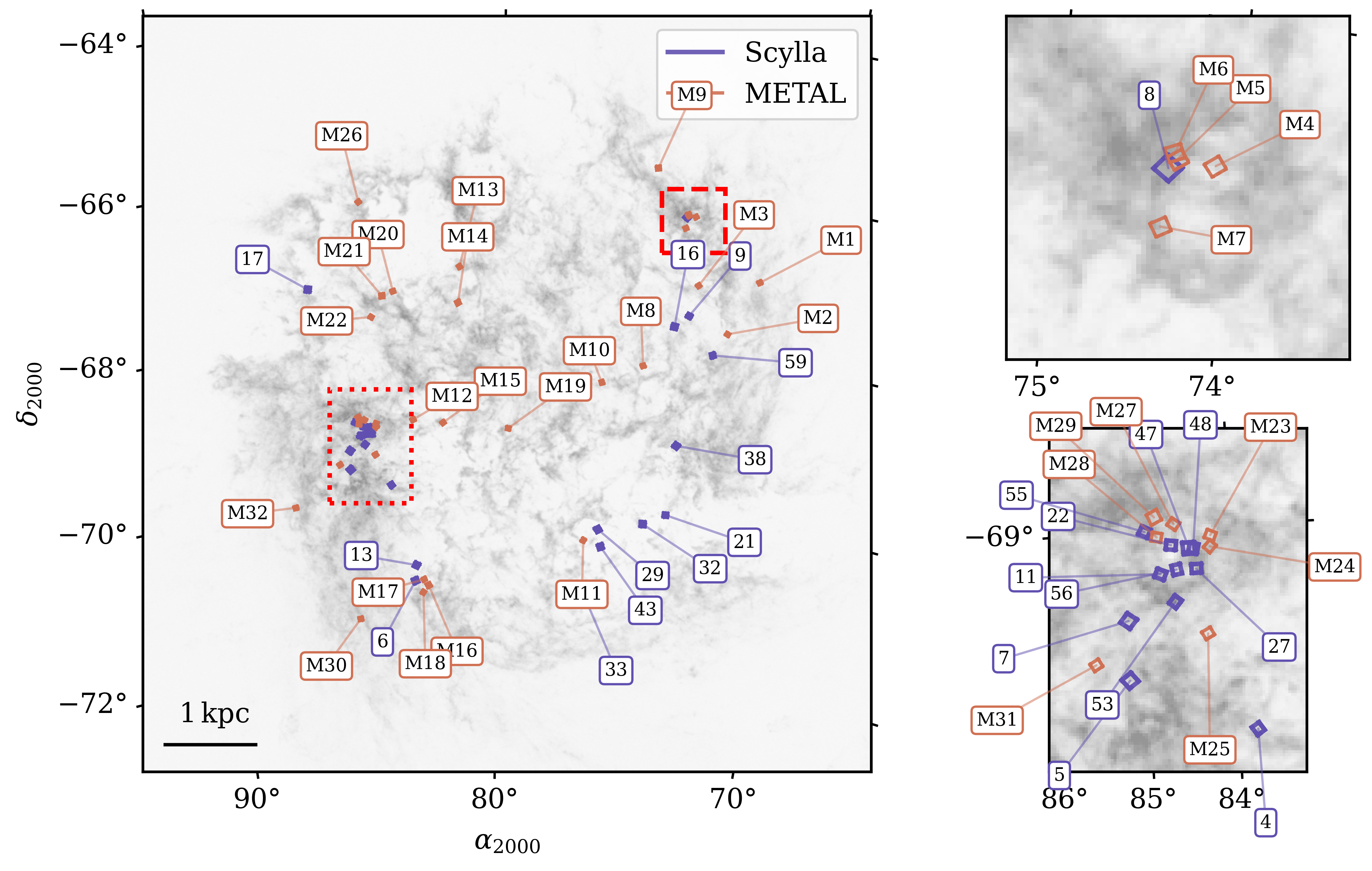}
	\caption{\textbf{Scylla and METAL field locations}: Maps of the peak brightness temperature of $21\rm\,cm$ emission in the SMC \citep[top;][]{pingel2022} and the LMC \citep[bottom;][]{kim1999}, overlaid with the footprints of the 57 Scylla ($N\geq3$ filters; blue) and 32 METAL fields (red). In each panel, regions where fields overlap are identified by dashed red bounding boxes and expanded at right. Each field is tagged by its ``short" name number where the METAL fields start with ``M" (see Table \ref{tab:fields}). At the distances of the LMC and SMC (50-60 kpc), the angular coverage of a single field ($160\times 160$ arcsec) translates to roughly 40-45 pc in physical scales.} 
    \label{fig:targets}
\end{figure*}

In this work, we present new stellar fits and dust measurements for over 1.5 million stars with multi-band SEDs in the MCs. In Section \ref{sec:data}, we briefly describe the photometric catalogs used. In Section \ref{sec:fitting}, we describe the BEAST analysis in detail, providing an overview of the stellar, dust, and observation models implemented in the fitting process, describe quality cuts for selecting high-reliability sources. In Section \ref{sec:validation}, we use simulations to validate the accuracy of the BEAST as function of filter coverage. In Section \ref{sec:results}, we present stellar and dust 
\startlongtable
\begin{deluxetable*}{llcccccc}
\tablecaption{\label{tab:fields} Field Summary}
\tablehead{
\colhead{Field Name} &  \colhead{Field Name} & \colhead{Field Center RA} & \colhead{Field Center Dec} & \colhead{$N_{\rm filters}$} & \colhead{Guard Flag} & \colhead{$N_{\rm stars}$} & \colhead{$N_{\rm stars}$}\\
\colhead{(short)} & \colhead{(long)} &  \colhead{($^{\circ}$)} & \colhead{($^{\circ}$)} & \colhead{} & \colhead{} & \colhead{(Total)} & \colhead{(HRC)} \\
\colhead{(1)} & \colhead{(2)} & \colhead{(3)} & \colhead{(4)} & \colhead{(5)} & \colhead{(6)} & \colhead{(7)} & \colhead{(8)} 
}
\startdata
SMC$\_ \rm $5  &  15891$\_ \rm$SMC-3514se-8584  & 16.4523 & -72.835 & 3 &  0  & 30816 & 12956  \\
SMC$\_ \rm $6  &  15891$\_ \rm$SMC-3956ne-9632  & 15.9954 & -72.1305 & 3 &  0  & 25835 & 9334  \\
SMC$\_ \rm $8  &  15891$\_ \rm$SMC-3955ne-9818  & 16.0039 & -72.1341 & 3 &  0  & 25299 & 8888  \\
SMC$\_ \rm $10  &  15891$\_ \rm$SMC-3149ne-12269  & 15.0793 & -72.1515 & 7 &  1  & 15319 & 5849  \\
SMC$\_ \rm $11  &  15891$\_ \rm$SMC-8743se-11371  & 21.4361 & -73.1212 & 3 &  0  & 4633 & 870  \\
SMC$\_ \rm $13  &  15891$\_ \rm$SMC-2983ne-12972  & 14.9313 & -72.1748 & 5 &  0  & 16782 & 5890  \\
SMC$\_ \rm $14  &  15891$\_ \rm$SMC-3669ne-13972  & 14.9314 & -71.9391 & 3 &  1  & 21776 & 6612  \\
SMC$\_ \rm $15  &  15891$\_ \rm$SMC-4292sw-13841  & 9.562 & -73.4013 & 7 &  1  & 15054 & 4752  \\
\enddata
\tablecomments{HRC stands for high-reliability catalog. The complete table is available in machine-readable format in the online journal.}

\end{deluxetable*}
\label{tab:fields}
\noindent  measurements for all sources in the SMC and LMC. In Section \ref{sec:discussion}, we discuss how these results compare with existing literature on stellar populations, star-formation histories, and alternative ISM tracers of the MCs. In Section \ref{sec:summary}, we summarize our findings.

\section{Data}\label{sec:data}

The Scylla survey is a pure-parallel program that obtained multi-band HST Wide Field Camera 3 (WFC3) imaging of 96 fields in the SMC and LMC during 
Cycles 27-31 \citep{murray2024b}. All fields were observed in at least two optical bands (F475W and F814W) and, if scheduling allowed, additional filters were included in the following priority order: F336W, F275W, F110W, F160W, F225W. UV filters were prioritized above IR because effects from dust extinction are strongest in the UV and measuring both sides of the Balmer jump (e.g. F336W combined with F475W) helps constrain stellar temperatures. To accurately measure both the stellar type and line-of-sight dust, we therefore only fit sources in fields with coverage in both the optical filters (F475W and F814W) and at least one UV filter, usually F336W, leaving us with 33 fields in the SMC and 24 fields in the LMC, the locations of which are shown in Figure \ref{fig:targets} (blue).   

In addition to the 57 Scylla fields with optical+UV coverage, we also include 32 fields from the Metal Evolution Transport and Abundance in the LMC (METAL) survey \citep{jrd2019}. METAL is a Cycle 24 HST (2016-2017) survey that obtained spectra towards 33 massive stars in the LMC while simultaneously obtaining parallel imaging using WFC3, similar to Scylla. Out of the 32 parallel WFC3 fields, 25 fields have seven-filter coverage with the same filters as in Scylla described above, one field has six-filter coverage, and six fields have four-filter coverage. The inclusion of these fields more than doubles the number of sources we can characterize in the LMC, the locations of which are shown in Figure \ref{fig:targets} (red).

\begin{figure*}
    \centering
    \includegraphics[width=0.76\linewidth]{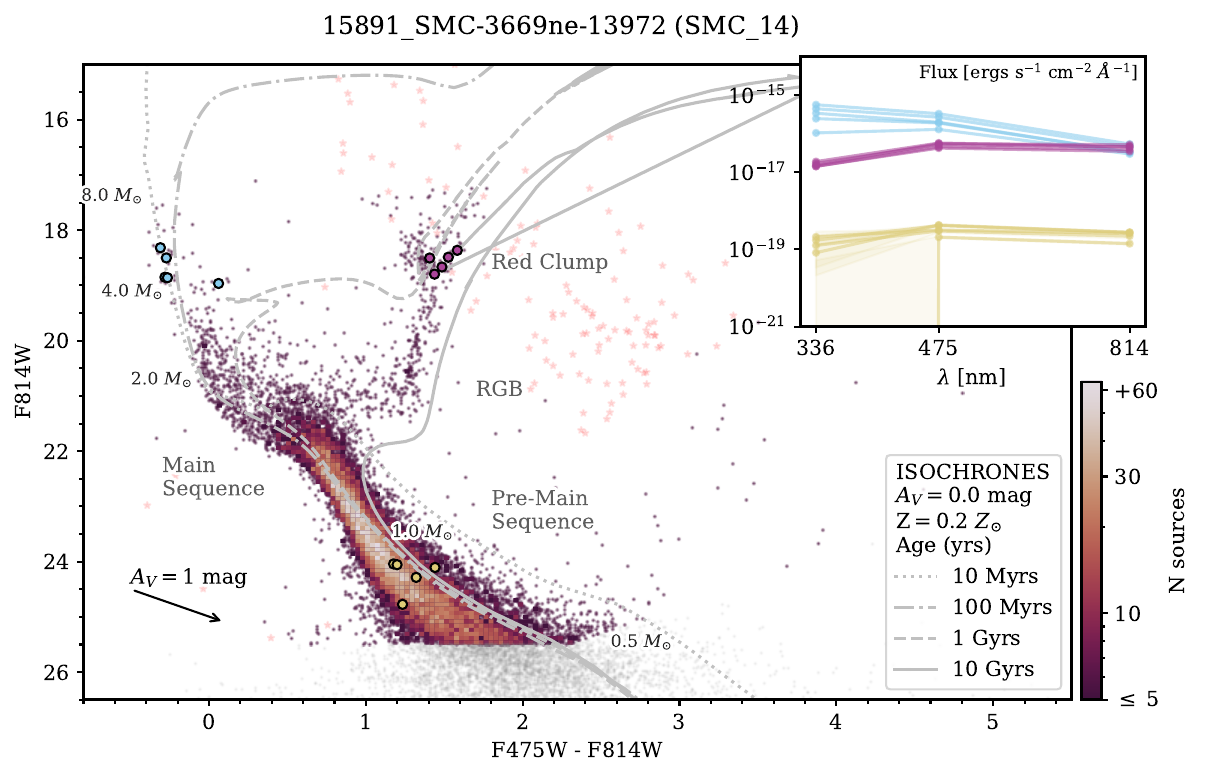}
    \includegraphics[width=0.76\linewidth]{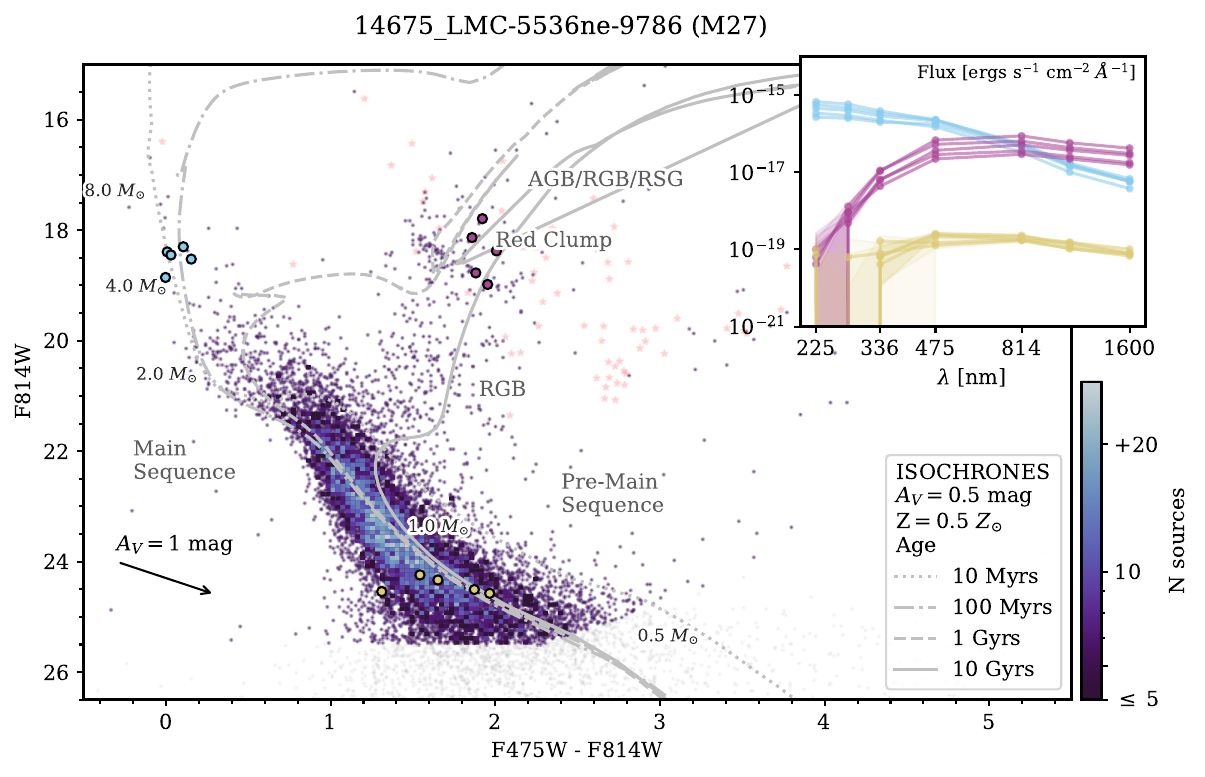}
    \caption{\textbf{Example CMDs}: Optical CMDs for one three-filter METAL field in the SMC (top; 15891{\textunderscore}SMC-3669ne-139720 or SMC{\textunderscore}14) and one seven-filter Scylla field in the LMC (bottom; 14675{\textunderscore}LMC-5536ne-9786 or M27). All sources below the 50\% completeness limits in either filter are shown in grey \citep[F475W=27.8 mag, F814W = 25.5 mag;][]{murray2024b}. Brighter sources (F814W $<$ 17 mag) are absent in the SMC field due to a lack of guard exposures that capture brighter sources before they saturate, independent of the number of filters observed. PARSEC isochrones are overlaid, ranging from 10 Myrs to 10 Gyrs. For the SMC field, we apply no extinction, assume a distance of 60 kpc, and $Z=0.2\ Z_{\odot}$, while for the LMC field, we apply 0.5 mag of extinction, assume a distance of 50 kpc, and $Z=0.5\ Z_{\odot}$. In both fields, we show the direction of a 1-magnitude reddening vector (black arrow). The location of potential foreground sources from TRILEGAL are plotted as faint red stars. \textbf{Example SEDS}: Inset plots (top right) show SEDs for five random sources in each of three distinct regions of the CMD: the upper main sequence (blue), the red clump (magenta), and the lower main sequence (yellow). Our ability to accurately model stars and their extinction is limited by the number of bands observed (i.e. more bands provide greater leverage).}
    \label{fig:eg_cmd}
\end{figure*}

In total, the data sample consists of 33 fields in the SMC and 56 fields in the LMC. Each field covers $160
\times 160$ arcsec (7.11 arcmin$^2$), except for fields with additional WFC3 IR coverage, in which case a field spans $123 \times 137$ arcsec (4.65 arcmin$^2$). At the distances of the LMC and SMC (50-60 kpc), the angular coverage of a single field translates to roughly 40-45 pc in physical scales.
Calibrated images for both METAL and Scylla were retrieved from the Mikulski Archive for Space Telescopes (MAST) archive and processed using CalWF3 versions 3.3 or 3.4.1 for METAL, and versions 3.5.0, 3.5.1, 3.5.2, 3.6.1 or 3.6.2 for Scylla, based on the date of observations \citep{romanduval2019, murray2024b}. For UVIS exposures, both surveys use images that have been corrected for charge transfer efficiency (CTE) effects. The photometry for both METAL and Scylla was performed using the DOLPHOT package \citep{dolphin2002, dolphin2016}, using a similar photometry pipeline as the PHAT survey \citep{williams2014, murray2024b}. 

For all fields, we only analyze sources that pass the vgst (``very good star") cuts \citep{murray2024b}, which require the following: (1) a source must be measured in all bands of a field (i.e. if a field was observed in three filters, then a source should have measurements in all three filters, regardless of the their quality); (2) a source must have a signal-to-noise ratio (SNR) greater than 4 \textit{in at least one observed band}; (3) to remove diffraction spikes and background galaxies, the squared sharpness, roundness, and crowding of a source must be equal to or less than 0.15, 0.6, and 0.2, respectively in optical bands F475W and F814W; and (4) a source cannot be flagged as anything other than \texttt{FLAG}=0 or \texttt{FLAG}=2 in DOLPHOT, where 0 indicates a ``good" PSF extraction, and 2 indicates too many bad or saturated pixels. We include sources with \texttt{FLAG}=2 because a significant fraction of sources in certain fields are associated with bright stars of interest. We apply these same cuts to the METAL data for consistency. In total, we fit 623,763 sources in the SMC and 949,831 sources in the LMC, roughly 30\% and 25\% of the number of sources in the original catalogs, respectively.

\subsection{Example Fields}

Sources cover a wide range of colors and magnitudes. In Figure \ref{fig:eg_cmd}, we plot the optical color-magnitude diagram (CMD) of a seven-filter METAL field in the LMC, (14675{\textunderscore}LMC-5536ne-9786 or M27\footnote{Long field names are formatted as ``[\texttt{PID}]{\textunderscore}[\texttt{galaxy}]$-$[\texttt{distance from galaxy center in arcsec}][\texttt{cardinal direction}]$-$[\texttt{telescope rotation angle in deg (V3PA)}]'', where \texttt{PID} is the HST program ID under which the observation was taken (e.g.~\texttt{14675{\textunderscore}LMC-5945ne-34377}). \texttt{PID} = \texttt{14675} denotes METAL fields, whereas all higher \texttt{PIDs} denote Scylla fields.}, top) and a three-filter Scylla field in the SMC (15891{\textunderscore}SMC-3669ne-13972 or SMC{\textunderscore}14, bottom). The locations of these fields were selected to represent the average Scylla and METAL fields, which will be preferentially located in the vicinity of star-forming regions (e.g.~30 Doradus for M27). Even though additional filters are available, both CMDs are plotted using F475W and F814W, since these (and F336W) are the default filters available in every Scylla and METAL field. We grey out all sources below the 50\% completeness limits in either filter (F475W=27.8 mag, F814W = 25.5 mag) to visualize the completeness limits calculated in \citet{murray2024b}. 

From the CMDs, we see there are brighter stars (F814W $< 17$ mag) in the LMC field compared to the SMC field. The lack of bright sources is a result of stars saturating during long exposures. For the majority of Scylla fields, we were only able to obtain standard long exposures due to scheduling restrictions as a parallel program. This means we were unable to obtain short ``guard" exposures like METAL did, which would have allowed us to extract the photometry of bright sources. Consequently, we expect to observe few bright stars in the 56\% of Scylla fields missing guard exposures.

Due to the large field of view of the MCs, we anticipate a small fraction of the sources in our fields to be contamination from low-mass stars in the Milky Way. For both fields, we plot the CMD location of potential foreground sources from TRILEGAL as faint red stars \citep{Groenewegen2002, girardi2005, girardi2007, Vanhollebeke2009}. Most foreground sources are both redder and brighter than the main sequence in each field, however, there is some overlap with red clump stars. Foreground sources will be poorly modeled at a distance of 50-60 kpc, so we will remove these sources with quality cuts based on modeling agreement (Section \ref{sec:quality}). 

In both fields, we observe a variety of CMD features such as a prominent main sequence, a track of red giant branch (RGB) stars, red clump (RC) stars, and pre-main sequence stars. In the LMC METAL field, we also observe even brighter RGB stars overlapping with a potential asymptotic giant branch (AGB) and red super giant (RSG) stars. This track is likely absent in the SMC field due to the brighter saturation limit. 

We overlay PARSEC isochrones to help us contextualize the range of masses and ages present in the fields. The isochrones span stellar ages of 10 Myrs (dotted) to 10 Gyrs (solid) \citep{bressan2012, chen2014, chen2015, tang2014, marigo2017}, assuming a distance of 60 kpc and 50 kpc and metallicities of 20\% and 50\% $Z_{\odot}$ for the SMC and LMC, respectively. In the SMC field, a 20\% $Z_{\odot}$ isochrone is well-matched with the observed upper main sequence, but is a poor fit to the RGB and red clump. Since RGB stars are several Gyrs older than upper main sequence stars, we expect RGB stars to be more metal-poor (e.g.~10\% $Z_{\odot}$). Previous Scylla papers report similarly low metallicities for RGB stars \citep[e.g.~Figure 2 in][]{cohen2024b}. RGB stars indicate the presence of older stellar populations, however, we also observe several sources in both fields that are likely younger than 100 Myrs, implying both galaxies have ongoing star formation, which is consistent with recent star formation history results \citep{mazzi2021, Jones2023, nayak2024}. For the 10 Myr isochrone (dotted), we note the CMD location of stars with initial masses ranging from 0.5 to 8 $M_{\odot}$. In the LMC field, there are a handful of sources that could potentially be more massive than 8 $M_{\odot}$. Based on the completeness limits, we should be able to detect stars with initial masses lower than 1.0 $M_{\odot}$. 

In the LMC field, we apply 0.5 mag of extinction to the isochrones to improve alignment with the observations. In general, we expect the LMC to contain more dust than the SMC, as evidenced by the width of the RGB and red clump. Even with this adjustment, the 10 Gyr isochrone is still bluer than a significant fraction of sources in the RGB ($\sim30$\%), a potential indication that these stars have more dust along their line-of-sight, as initially shown in Figure 15 of \citet{murray2024b}.  Compared to the total internal extinction, we anticipate limited amounts of foreground extinction for both fields with an estimated $A_V = 0.1\mbox{--}0.2$ mags for the SMC and $A_V = 0.2\mbox{--}0.5$ mags in the LMC \citep{Schwering1988, staveleysmith1997, chen2022}. 
 
Although both fields have similar optical CMDs, our ability to characterize individual SEDs is highly dependent on the number and range of filters observed. In the inset plots (Figure \ref{fig:eg_cmd}, top right), we show the SEDs for five random sources in three separate regions: (1) the upper main sequence (blue); the RGB/RC (magenta); and the lower main sequence (yellow). The CMD locations of these sources are also indicated in the main panel. For the seven-filter LMC field, we see that the upper main sequence stars (blue) peak in the UV, while the RGB/RC stars (magenta) are also bright but peak in the optical bands and drop off sharply in the UV. Meanwhile, lower main sequence stars (yellow) are several orders of magnitude dimmer in all bands and have much larger photometric uncertainties (shaded). For the three-filter SEDs in the SMC field, we observe similar fluxes, however, the overall shapes of the SEDs are much less apparent. Therefore, we anticipate that SED parameter fits for sources in fields with fewer filters (e.g.~3 bands) will be less constrained than fields with more filters (e.g.~+4 bands). We investigate the effects of filter number on BEAST constraints in Section \ref{sec:validation}.

Going forward, we will refer to the Scylla and METAL fields analyzed in this paper collectively as Scylla fields for simplicity. 

\section{Analysis}
\label{sec:fitting}

\begin{deluxetable*}{lccccccc}[ht]
\tablecaption{BEAST Model Parameters}\label{tab:beast}
\tablehead{
\textbf{Parameter} & \textbf{Unit} & \textbf{Description} & \textbf{Min} 
& \textbf{Max} & \textbf{Resolution} & \textbf{Prior} }
\startdata
$\log(A)$ & years & stellar age & 6.0 & 10.13 & 0.1 & flat SFR \\
\textbf{$\log(M_{ini})$} & $M_{\odot}$ & \textbf{initial} stellar mass & $-0.8$ & 2.0 & variable$^a$ & Kroupa IMF \\
$\log(Z)$ & $Z_{\odot}$ & stellar metallicity & $-2.1$ & $-0.3$ & 0.3 & flat \\
$A_V$ & mag & dust column & 0.01 & 10.0 & 0.05 & flat \\
$R_V$ & $\cdots$ & dust average grain size & 2.0 & 6.0 & 0.5 & peaked at $\sim$2.7$^b$ \\
$f_\mathcal{A}$ & $\cdots$ & dust mixture coefficient & 0.0 & 1.0 & 0.1 & peaked at 1$^b$ \\
$d$ [LMC] & kpc & distance & 40 & 60 & 2.5 & flat \\
$d$ [SMC] & kpc & distance & 47 & 77 & 2.5 & flat \\
\hline
\enddata
\centering
\tablecomments{$a$: Initial stellar mass resolution is variable and depends on sampling of stellar evolutionary models.\\
$b$: $R_V$ and $f_\mathcal{A}$ priors are set by 2D prior, described in Fig.~7 of \citet{gordon2016}. Weighted average of the 2D prior is $R_V=3.55$ and $f_\mathcal{A}=0.73$.}
\end{deluxetable*}

We use the Bayesian Extinction And Stellar Tool \citep[BEAST;][]{gordon2016} to analyze the Scylla photometric catalogs. The BEAST uses a Bayesian probabilistic approach to model the intrinsic properties of individual stars and dust along each sightline based on their multi-wavelength spectral energy distributions (SEDs). 
While the BEAST is adept at analyzing individual stars, it is optimized for large-scale photometric surveys of resolved stellar populations, where factors like crowding and photon noise significantly influence flux measurements and photometric biases across filters. The BEAST is also designed to incorporate priors to accommodate previous knowledge about the expected stellar and dust properties from relevant studies (e.g.~mass distribution/initial mass function, age distribution/star-formation history, distance ranges, etc.). 
The BEAST fitting procedure is explained in detail in \citet{gordon2016}, where the BEAST was used to interpret observations from the Panchromatic Hubble Andromeda Treasury \citep{dalcanton2012}. Since then, the BEAST has been applied to other galaxies in the Local Volume and is established as a reliable SED fitting tool \citep{vandeputte2020, choi2020, yanchulova2024}.

In Section \ref{sec:fitoverview}, we provide a brief summary of the BEAST grid construction and the stellar, dust, and observational models employed for version 1.0 of the Scylla fitting. Details regarding the range, resolution, and priors for each parameter in the grid are summarized in Table~\ref{tab:beast}. In Section \ref{sec:fit}, we describe the summary statistics the BEAST provides from its fitting procedure and present an example fit towards two stars: one with noticeable extinction and one without (Figure \ref{fig:sed_example}). In Sections \ref{sec:quality}, we present quality cuts to isolate high-reliability sources.

\subsection{Predicted SEDs}
\label{sec:fitoverview}

To characterize individual stars, the BEAST compares the observed SED of each star with predicted SEDs. These predicted SEDs are constructed using a range of inputs from the stellar and dust models with corresponding priors, producing a seven-dimension grid (one for each parameter; see Table~\ref{tab:beast}) of extinguished SEDs to which we can compare the data. 
To construct this grid, first, the BEAST generates a grid of stellar spectra across a range of initial masses, ages, and metallicities (details in Section~\ref{sec:stellarmodel}). Then, varying amounts of extinction are applied to each stellar spectrum based on the dust extinction models (defined by three dust parameters detailed in Section~\ref{sec:dustmodel}), producing an even larger grid of \textit{extinguished} spectra. The grid is then duplicated at a range of distances, creating a seven-dimensional parameter grid of extinguished stellar spectra. We then integrate these spectra over $HST$ transmission curves for each observed filter to generate model photometric measurements (i.e. simulated SEDs) for every combination of input stellar and dust parameters. 

Lastly, to account for systemic observational effects in the actual photometry, we use artificial star tests to quantify the average photometric bias as a function of flux in each filter (details in Section \ref{sec:noisemodel}). These biases are added to the predicted SEDs so we can accurately compare with the observed SEDs.

We note that the number of data points does not restrict the number of model parameters that can be fit since the BEAST is not fitting SEDs with a set of basis functions that can describe any shape (e.g.~polynomials). Rather, we are fitting with a limited set of shapes (i.e., dust extinguished stellar SEDs), which means that even with seven stellar and dust parameters, there are SED shapes that cannot be fit well (e.g.~unresolved binaries, stars with complex atmosphere models like AGB stars, non-stellar emission sources, etc.). Additionally, certain parameters (e.g.~mass, age, extinction) will have a much larger influence on the resulting SED shapes than other (e.g.~stellar metallicity, dust composition, etc.).

\subsubsection{Stellar Model}
\label{sec:stellarmodel}

The stellar model SED grid is constructed by integrating stellar evolutionary and atmospheric models. The primary stellar parameters consist of the initial stellar mass ($M_{\rm ini}$), age ($A$), and metallicity ($Z$), from which we can derive stellar parameters such as the stellar effective temperature ($T_{\rm eff}$), surface gravity ($g$), and luminosity ($L$). For the stellar evolutionary models, we use PARSEC stellar evolutionary tracks sourced from the \texttt{CMD 3.7} web interface\footnote{http://stev.oapd.inaf.it/cgi-bin/cmd} of \cite{bressan2012}, assuming a \cite{kroupa2001} initial mass function (IMF), and encompassing a metallicity range approximately spanning from 0.008 to 0.5 Z$_{\odot}$ \citep{lee2005}. 
We use stellar atmosphere grids developed by \citet{castellikurucz2003} and \citet{lanzhubeny2003, lanzhubeny2007} for local thermodynamic equilibrium (LTE) models and non-LTE models, respectively. Where the two grids overlapped, we preferred the \citet{lanzhubeny2003, lanzhubeny2007} models \citep[see Section 3.1 in][]{gordon2016}.


It is important to acknowledge that the BEAST fitting process has recognized limitations as it presently lacks models for certain stellar types including binaries, AGB stars, young stellar objects, and other rare stellar species. These stellar types require complex modeling and are therefore not usually included in common stellar evolutionary models. These omissions will impact the accuracy of fits for any unrepresented stellar types. For example, AGB stars are often characterized as highly extinguished red super giants since these two stellar types are relatively co-spatial in a CMD \citep{lindberg2024}. In addition, unresolved binaries will often appear redder and/or more luminous than a singular primary star due to the presence of a lower-mass secondary. However, without binary models, the BEAST will attempt to model these stars by applying increasing amounts of dust, resulting in an overestimation of extinction (e.g.~Section \ref{sec:binaries}). As a check, we can place upper limits on the expected extinction by observing the maximum extinction of brighter and older stellar populations, such as RGB stars \citep{dalcanton2023}, which generally have a scale height larger than star-forming thin disks of galaxies \citep{patra2019, meng2021}. For most of these unrepresented stellar types, there are often few SED models that can reproduce the observed magnitudes and colors, meaning that uncertainties on parameter fits will be unusually small relative to other common stellar types. This quirk gives us an opportunity to remove misfit stellar types by analyzing the uncertainties of the BEAST results, rather than just relying on the photometry.

\subsubsection{Dust Model}
\label{sec:dustmodel}

One of the main science objectives of the BEAST is to quantify not only the amount of dust, but also how the nature of dust grains varies within galaxies. For most sightlines in the Milky Way (MW), variations in the shape of the extinction curve can be parameterized as a function of $R_V$, the ratio of total extinction, $A_V$, to selective extinction, $E(B-V)$. $R_V$ reflects the grain size distribution \citep{cardelli1989, fitzpatrick1999, gordon2023}. However, the majority of extinction curves measured towards the MCs do not display this dependence on $R_V$ \citep{gordon2003, gordon2024}. Instead, notable distinctions are observed in the measured SMC extinction curves, particularly in the UV part of the spectrum, where the ``SMC Average" curve exhibits the absence of a 2175~\AA\ bump and a steeper rise in the far-UV. These variations are often attributed to the lack of metals in the MCs, but the large extinction curve variations within the MCs indicate other effects are important and may even be dominant \citep{misselt1999, gordon2003, gordon2024}.  Regardless, extinction curves within the Local Group exhibit a wide range of properties falling between those of the MW and the SMC extinction curves \citep{gordon2003, clayton2015, fitzpatrick2019, gordon2024}.

To address this variability, \citet{gordon2016} proposed a mixture model for dust extinction where the coefficient $f_\mathcal{A}$ represents the proportion of MW-type dust extinction and 1 - $f_\mathcal{A}$ represents the proportion of SMC-type dust extinction ($A_{\lambda}$ signifies the extinction at wavelength $\lambda$):

\begin{equation}\label{eqn:eqn_mixture_dust} 
\frac{A_{\lambda}}{A_V} = f_\mathcal{A} \left[\frac{A_{\lambda}}{A_V} \right]_{\mathcal{MW}} + (1 - f_\mathcal{A})\left[\frac{A_{\lambda}}{A_V} \right]_{\mathcal{SMC}}
\end{equation}

\noindent where $[A_\lambda/A_V]_{MW}$ and $[A_\lambda/A_V]_{SMC}$ are the average MW and SMC curves.
The mixture model smoothly interpolates between SMC-type dust \citep[$R_V$ = 2.74; $f_\mathcal{A}$ $\sim$ 0; ][]{gordon2003} and MW-type diffuse ISM dust \citep[$R_V$ = 3.1; $f_\mathcal{A}$ $\sim$ 1; updated to][]{fitzpatrick2019}. Future work will use the updated SMC Average curve \citep{gordon2024}.

We analyze the Scylla data with this flexible dust model using three parameters: $A_V$, the total extinction in V-band; $R_V$, the total-to-selective extinction; and $f_\mathcal{A}$, a mixture coefficient indicating the proportion of MW- and SMC Bar-type dust \citep[e.g., range of 2175~\AA\ bump strengths;][]{gordon2016}. BEAST extinction measurements will consist of both internal galactic extinction from the SMC/LMC and foreground extinction from the Milky Way because the BEAST uses individual stars as sightlines. Due to this foreground extinction, we expect the minimum potential extinction for extragalactic sources to be greater than zero. Other dust measuring techniques, such as quantifying the reddening vector of RGB stars on a CMD, only measure the internal extinction of a galaxy \citep{dalcanton2015, ymj2017}. 

Most dust column densities tend to exhibit power-law or log-normal distributions \citep{dalcanton2015, Lombardi2015, Schneider2015}, with a tail of higher densities. Given the vast range of star-forming environments observed across the Scylla fields and the fact that we are probing the dust column densities with discrete sightlines, we want to ensure that we do not unintentionally bias BEAST fits against higher extinction. As such, we choose to adopt a flat prior for $A_V$. Meanwhile, for $R_V$ and $f_\mathcal{A}$, we adopt the same 2D prior as described in \citet{gordon2016}, which have 1D projections that peak at $\sim3$ and 1, respectively.

\begin{figure*}
    \centering
    \includegraphics[width=0.85\textwidth]{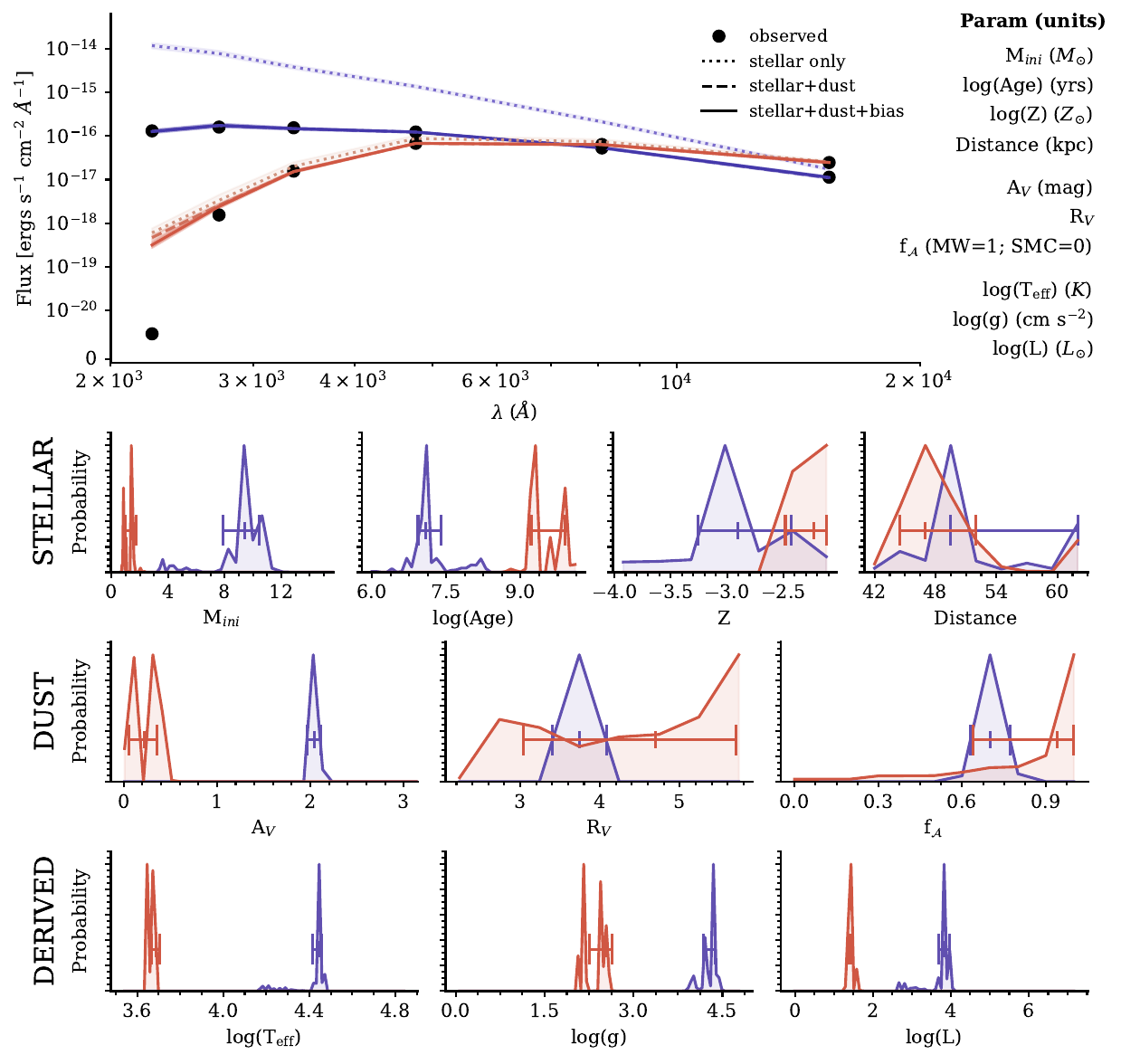}
    \caption{\textbf{Example SED fits from the BEAST}: SED fits of two stars in an LMC field (16786{\textunderscore}LMC-5127ne-3118) - a massive star with moderate amounts of extinction (blue) and a red giant branch star with almost no extinction (red). The observed fluxes are shown with black dots. The best-fit unextinguished stellar SED (dotted), extinguished SED (dashed), and bias-corrected SEDs (solid) are shown for comparison (dashed and solid lines are almost entirely overlapping). Model SEDs are constructed using the median ($\pm1\sigma$) values of each marginalized stellar and dust parameter. In the panels below, we show the distributions of these 1D marginalized parameters, noting the median $\pm1\sigma$ values. For both stars, there are several combinations of stellar and dust parameter values that can produce the observed flux. Primary parameters such as initial mass ($M_{ini}$), age, and extinction ($A_V$) are more tightly constrained compared to secondary parameters like metallicity ($Z$), distance, total-to-selective extinction ($R_V$), and extinction curve shape ($f_\mathcal{A}$) which have a weaker effect on the observed flux.
    }    
    \label{fig:sed_example}
\end{figure*}

\begin{figure*}
    \centering
    \includegraphics[width=0.85\textwidth]{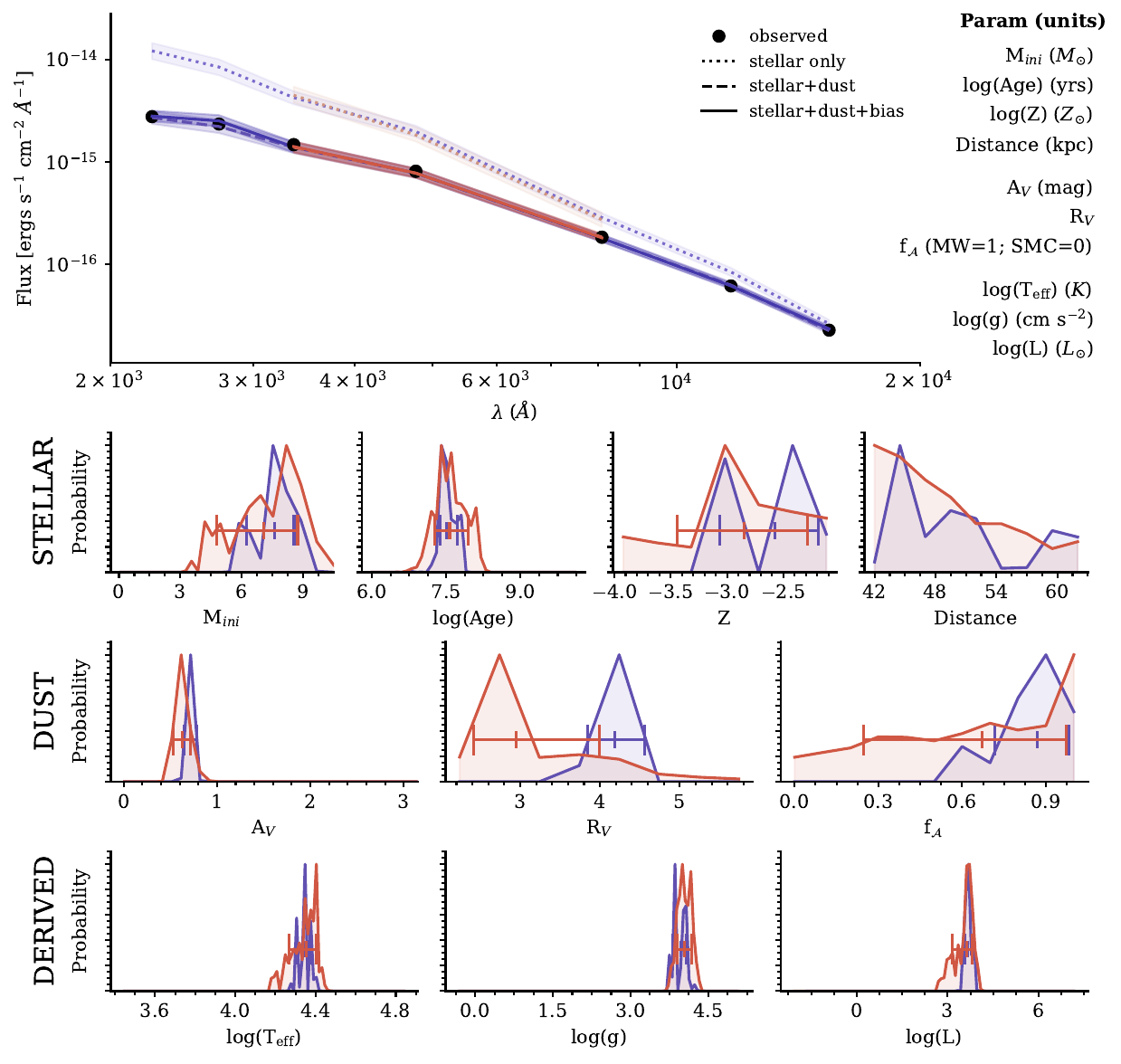}
    \caption{\textbf{Example SED fit from the BEAST with differing filter coverage (same as Figure \ref{fig:sed_example})}: SED fit of one stars in an LMC field (14675{\textunderscore}LMC-15562nw-33974) showing how BEAST parameters change when fit with three optical filters (red) versus seven NUV-NIR filters (blue). With additional filter coverage, we can improve uncertainties on parameters like mass ($M_{ini}$) and age ($\log(Age)$), and reduce biases in parameters like extinction ($A_V$), metallicity ($Z$).}   
    \label{fig:sed_example_filters}
\end{figure*}

\subsubsection{Observational Model}
\label{sec:noisemodel}

One of the main issues when comparing the observed photometry of stars to modeled stellar photometry is that we do not know how observing conditions might influence observations. Factors like extended background emission and nearby sources (e.g.~binaries, background galaxies, other stars) can all bias what fluxes we observe for an individual source, or whether we even detect a source at all. In particular, stellar crowding due to neighboring sources can result in contaminated flux measurements, causing independent flux offsets in each filter. These effects can be significant in fields with high source densities, such as the center of M31, which are considered crowding-limited in \textit{HST} data when stellar densities exceed 10 sources/arcsec$^2$ \citep[see Figure 33 in][]{williams2014}. We anticipate the impact of stellar crowding in Scylla fields will be minimal due to the relative proximity of the MCs -- with most Scylla fields having stellar densities $<1$ sources/arcsec$^2$ \citep{murray2024b}. However, quantifying the observational effects in each Scylla field it is still necessary if we want to compare observations with model SEDs. 

To address these issues, we develop an observational model to simulate observational effects in the data. Specifically, the observational model quantifies the average bias between model SEDs and their simulated observations and average observational uncertainty across a range of flux bins for each filter. To simulate observations, we employ artificial star tests (ASTs) using the following system: (1) photometry is simulated for a large number of stars across a range of fluxes in each observed filter (usually several thousand); (2) for each star, its intrinsic flux is converted to electron counts which are added to the original observations of a field at a random location within the images; and (3) observations are re-reduced using the same photometry pipeline as the original observations. Using this process, we can determine whether a star that was injected into the image was recovered or not, helping us quantify what fraction of stars we might be missing at each magnitude (i.e. how ``complete`` our photometry is at increasing magnitudes). If a star is observed, we can then measure the photometric bias, i.e.~how much its observed flux changed compared to its injected flux, and the photometric uncertainty, i.e. the standard deviation of the measured fluxes. Using this information, we calculate the average bias and uncertainty of sources in each filter as a function of the observed flux. A summary of the observational model can be found in Section 4.4 of \citet{murray2024b}.

We require ASTs to cover a wide range of fluxes to ensure that we are able to quantify observational effects for any realistic stellar model (realistic in this case means having at least one band within the range of observed fluxes). To do this, we estimated the total flux range in each observed band and split that range into 40 flux bins. Model SEDs are then randomly selected from the BEAST predicted SEDs until there are at least 50 model SEDs in each bin, resulting in a few thousand SEDs per field. By selecting stars in this manner, we assume bands are independent. Realistically though, we do expect correlation between observed bands, however, quantifying these correlations would require far more ASTs. In other words, fewer samples are needed to cover $N$ 1-dimensional spaces, where $N$ is the number of observed bands, as opposed to one $N$-dimensional space. By assuming the bands are independent, we thus minimize the number of ASTs needed to construct the observational model.

\subsection{Fitting SEDs}\label{sec:fit}

For each star, we quantify the likelihood of each model by directly comparing the synthetic photometry with the observed photometry and use the photometric uncertainties from the observational model as the uncertainties in the calculations, producing a grid of likelihoods. This likelihood grid is multiplied by a similarly-sized grid of priors, providing 7D posterior probability distribution functions (pPDFs) for all relevant models. We marginalize this 7D pPDF to obtain 1D pPDFs for each input parameter. 

The BEAST reports a sparse sampling of the full 7D pPDFs, the 1D marginalized pPDFs themselves, and 16th, 50th, and 84th percentile values from the 1D pPDFs (i.e.~median $\pm 1 \sigma$) for each parameter. The BEAST also reports the expectation value (\texttt{Exp}) -- the weighted average of the posterior, and the parameter values for the single best model fit from the whole grid (\texttt{Best}) along with the corresponding $\chi^2$ of said \texttt{Best} fit. 

In Figure \ref{fig:sed_example}, we show example SED fits for two stars in the catalog, one with significant extinction and one without. In the main upper panels, the black dots show the observed flux of each star across the six observed filters and overlaid in shaded colors are the simulated SEDs based on the median $\pm 1 \sigma$ fits of the marginalized parameters. In the panels below, we plot the marginalized 1D pPDFs for each BEAST parameter, and note the 16th, 50th, and 84th percentile fits based on the 1D pPDFs. 
Going forward, all parameter measurements ($X$) will be based on the median values of their 1D pPDFs ($X_{p50}$), and parameter uncertainties will be based on the 1$\sigma$ deviations defined as: ($X_{p84} - X_{p16}$)$/2$.

\begin{figure*}[!t]
    \centering
    \includegraphics[width=\textwidth]{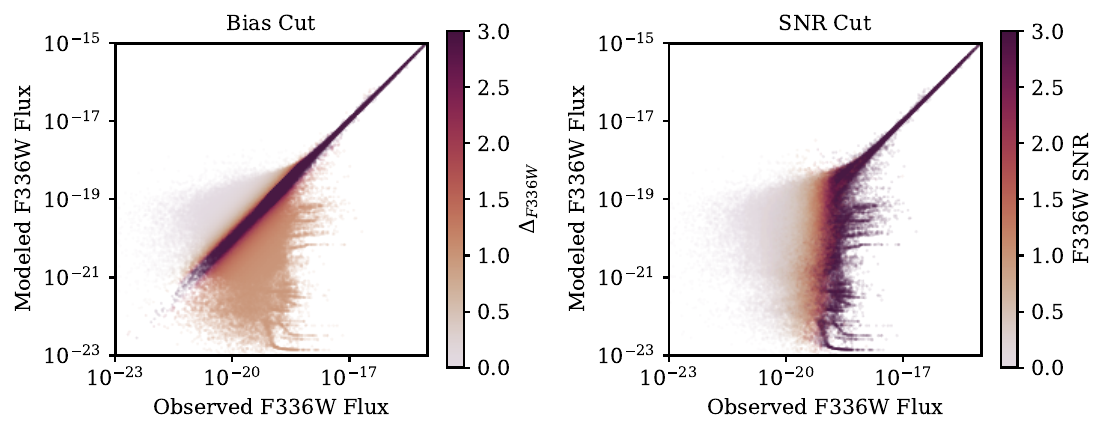}
    \caption{\textbf{High-reliability quality cuts}: Left: Observed F336W flux versus the BEAST-modeled F336W flux of all sources in the SMC and LMC, colored by their signal-to-bias ratio ($\Delta_{\mathrm{F336W}}$). Right: Same as left but sources are colored by their photometric signal-to-noise ratio in F336W. For visual clarity, we limit the color bars to 3.0 for both plots. We exclude any sources with $\Delta_{\mathrm{F336W}} < 3$ \textit{or} SNR$_{\mathrm{F336W}}< 3$ from the high-reliability catalogs.
    }
    \label{fig:quality_cuts}
\end{figure*}

From the marginalized 1D pPDFs, we see there several parameters that exhibit double peaked probability distribution, specifically $M_{ini}$, age, and $A_V$. In these scenarios, there are multiple combinations of parameter values that could produce SEDs similar to the observations. Bimodalities in the parameter fits arise when there are observational degeneracies between luminosity, distance, and extinction. However, for most 1D pPDFs, there is generally only one major mode which is often aligned with the median, and even if there are multiple modes of similar probability (e.g.~$A_V$ for the red star), these modes tend to be clustered in the same region of the parameter range.

Not all Scylla fields were observed with the same number of filters. In Figure \ref{fig:sed_example_filters}, we show how BEAST fits for a single star varies when observed with seven NUV-NIR filters (blue) versus three optical filters (red). While both  extinguished stellar model SEDs agreed with the observed photometry (black dots), in the parameter subfigures, we see notable differences in the marginalized 1D pPDFs. Primary parameters, like initial mass ($M_{ini}$), age, and extinction ($A_V$), were all recovered in a similar range, although the seven-filter BEAST fits have smaller uncertainties. Meanwhile, secondary parameters like metallicity ($Z$), distance, $R_V$, and $f_\mathcal{A}$ all span a wide range of values and are generally in poor agreement. In Section \ref{sec:validation}, we investigate this effect of filter coverage in more detail. 

\subsubsection{Fitting Procedure}

To perform the fitting, we made use of high-performance computing (HPC) resources of the Advanced Cyberinfrastructure Coordination Ecosystem: Services \& Support supercomputer program supported by the National Science Foundation. The resources were allocated on the Bridges-2 HPC platform administered by the Pittsburgh Supercomputer Center (PSC). The total resources used amounted to 4.5 million service units (SUs) on Bridges-2 Regular Memory (RM) nodes, and 220,000 SUs on the Bridges-2 Extreme Memory (EM) nodes. EM nodes were used for tasks that required a large amount of memory (e.g.~generating BEAST model grids and observational models, and trimming the BEAST model grids), while the RM nodes were used for general-purpose computing such as the fitting of each individual source. These resources were also used for overhead tasks, such as code benchmarking and occasional re-fitting.

Overall, we fit 949,831 stars in the LMC and 623,763 stars in the SMC, located across 56 and 33 individual fields in the LMC and SMC, respectively. To produce the final Scylla catalogs presented in this paper, we trim the BEAST parameter files consisting of fits for each individual star by post-processing them with the quality cuts described in the next section.

\subsection{High-Reliability Catalog Quality Cuts}\label{sec:quality}

Quality cuts in the Scylla photometric catalog were less restrictive compared to other photometric surveys \citep[e.g.~][]{dalcanton2012, williams2014}, with only single-band cuts on signal-to-noise ratios (SNR) and no restrictions on crowding (for more details, see Section \ref{sec:data}). This was done to ensure that as broad a range of potential sources were characterized by the BEAST and that no rare or interesting objects were excluded.
With BEAST fits for all of these sources, we now define additional quality cuts to establish a high-reliability catalog (HRC) flag to distinguish between well-fit sources and poorly-fit sources/potential contaminants.

BEAST likelihoods were calculated using all available wavelength information in each field. However, some Scylla fields fit with the BEAST only have 3-filter coverage, so we want to ensure that any quality cuts can be applied across all fields equally. 
Between the optical and the UV filters shared in all fields (F814W, F475W, and F336W), the UV is the best band for determining whether extinguished models are actually doing an adequate job of matching the observed data. This is likely due to the fact that the majority of sources peak at optical wavelengths, and so the UV provides the first constraint on the overall shape of the SED, especially since F336W covers the Balmer jump. Almost all sources fit with the BEAST have UV coverage in F336W (the only exception being sources in field 14675\_LMC-5378se-19128, where we substitute F275W for F336W for all quality cuts), so we quantify how well the modeled UV flux matches the observed UV flux by calculating the difference between the two. Fainter sources will have smaller differences than brighter sources, so we normalize the difference by dividing by the observed flux. This fraction gives us a ratio between the observation and the bias, which we define as follows:

\begin{equation}
\Delta_{\theta} = \left(\frac{F(\theta)_\mathrm{Obs} - F(\theta)_\mathrm{Model}}{F(\theta)_\mathrm{Obs}}\right)^{-1} \label{eq:F336W_delta}
\end{equation}

with $\theta$ being F336W.

In Figure \ref{fig:quality_cuts} (left), we plot the observed F336W flux versus the BEAST predicted F336W flux of all sources, colored by the signal-to-bias ratio ($\Delta_{\mathrm{F336W}}$). We find that $\Delta_{\mathrm{F336W}} > 3$ manages to retain most bright sources while simultaneously removing sources with poor agreement (i.e. the bias between the modeled flux and the observed flux cannot constitute more than 1/3 of the observed flux).

\begin{figure*}
    \centering
    \includegraphics[width=0.87\textwidth]{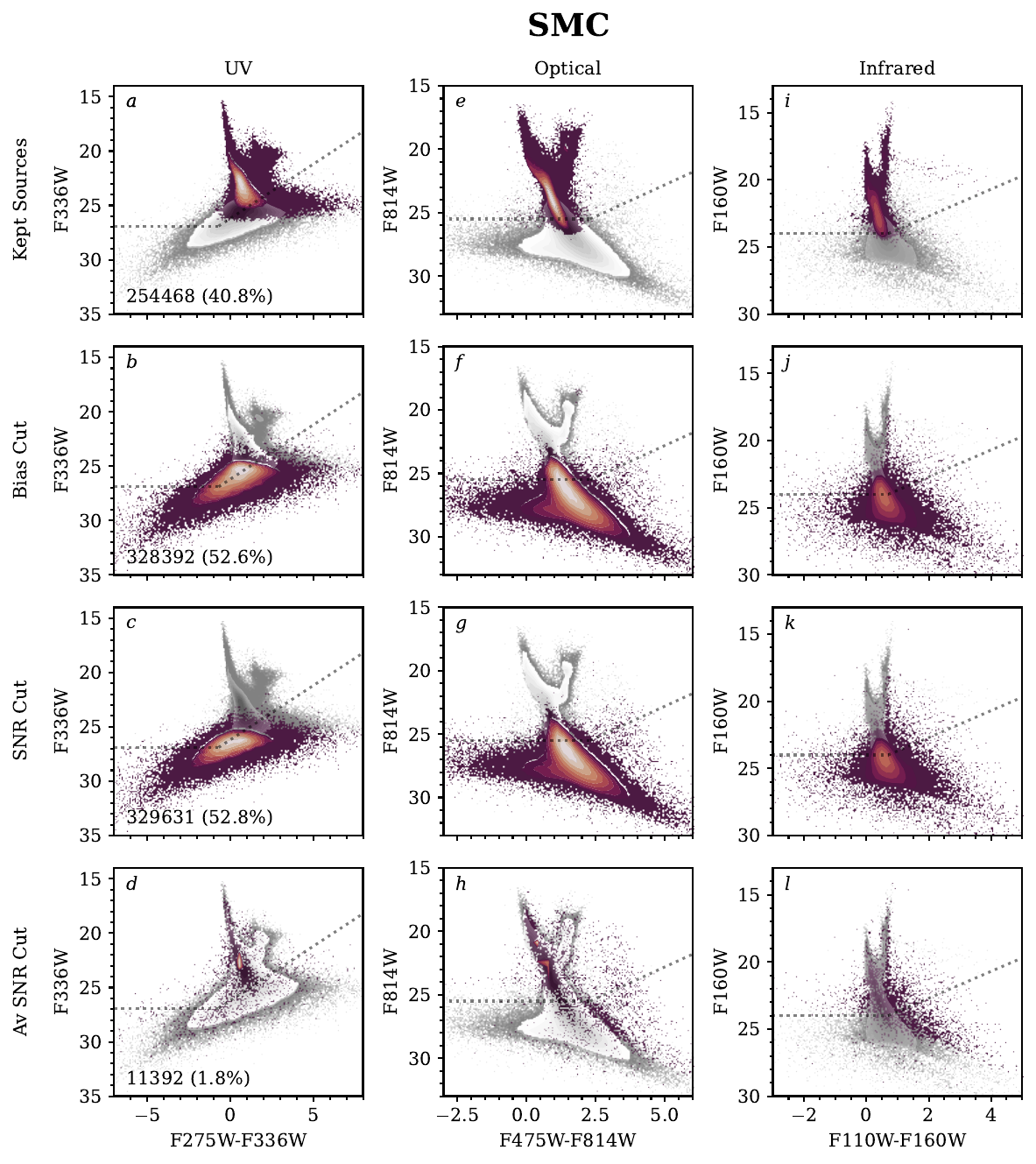}
    \caption{\textbf{High-reliability catalog quality cuts}: UV (left), optical (middle), and infrared (right) CMDs of all SMC sources that pass the HRC quality cuts (top rows), and sources that failed (lower rows). The inverted catalogs are shown in grey for reference. The majority of poorly-fit BEAST sources are dim and generally fall below the 50\% completeness limits calculated from artificial star tests (dotted lines), except for the $A_V$ SNR cut which targets brighter diffraction spikes.}
    \label{fig:quality_cuts_cmd_smc}
\end{figure*}

\begin{figure*}
    \centering
    \includegraphics[width=0.87\textwidth]{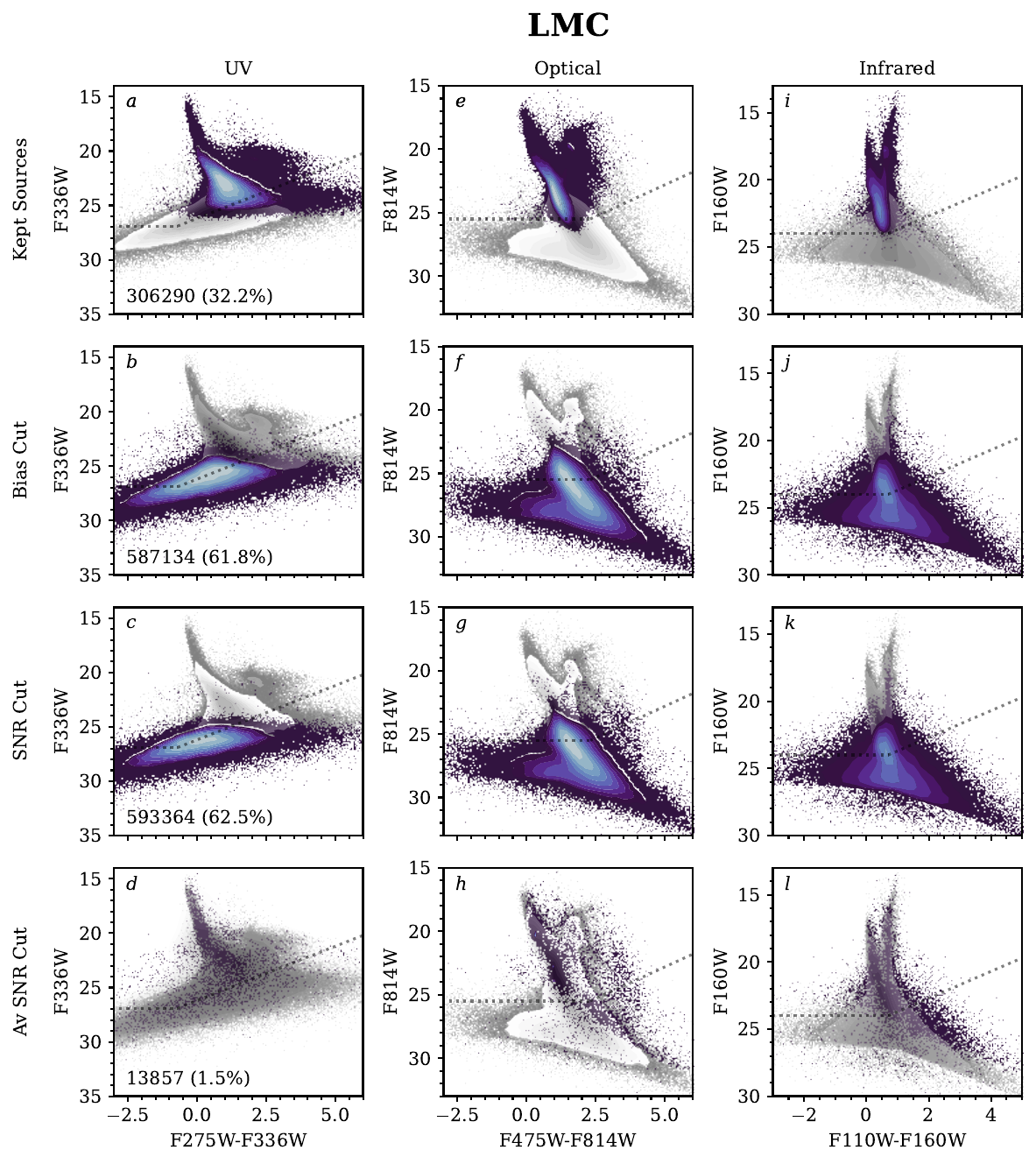}
    \caption{Same as Figure \ref{fig:quality_cuts_cmd_smc} but for the LMC: UV (left), optical (middle), and infrared (right) CMDs of all LMC sources that pass the HRC quality cuts (first rows), and sources that failed (lower rows). The inverted catalogs are shown in grey for reference.}
    \label{fig:quality_cuts_cmd_lmc}
\end{figure*}

We impose two additional quality cuts on sources for the HRC catalog. First, we require sources to have photometric SNRs $> 3$ in F336W since this is the only filter blueward of the Balmer jump available for all sources in every field (Figure \ref{fig:quality_cuts}; right). This removes ultra-faint sources below the 50\% completeness limits, as detailed in Section 4.4 of \citet{murray2024b}. While we originally included sources with low flux SNRs in the BEAST fitting process, we find that their resulting BEAST fits are generally poorly constrained, and we therefore chose to exclude them from the high-reliability catalog, however, BEAST fits for these sources are still available in the complete catalogs.

Second, although the vast majority of diffraction spikes were removed from the original photometry catalogs using photometric quality cuts \citep[see Appendix B in][]{murray2024b}, we typically find a handful of BEAST sources (less than a dozen) per field with abnormally high $A_V$ estimates and low $A_V$ uncertainties spatially coinciding with saturated sources in the images.
These sources have $A_V$ SNRs ($\mathrm{SNR}(A_V) = 2\times A_{V p50}/(A_{V p84}-A_{Vp16})$) much higher than all other sources in the catalogs: diffraction spike sources have $A_V$ SNRs from $10\mbox{--}20$, while all other catalog sources have $A_V$ SNRs $1\mbox{--}5$. Removing the top 1\% of high $A_V$ SNRs sources removes all potential contamination from diffraction spikes, except for in one field (16786\_SMC-4451ne-16362) where we opt to place the $A_V$ SNR manually at 10. We remove the top 1\% of high $A_V$ SNR sources for each field after placing the previous two quality cuts. By isolating potential diffraction spikes via an SNR cut rather than a cut on $A_V$ directly, we are able to retain the majority of high $A_V$ sources in each field.

Cumulatively across all fields in the SMC and LMC, these quality cuts remove roughly 59\% and 68\% of sources from the BEAST catalogs. The final high-reliability catalogs contain 254,468 and 306,290 in the SMC and LMC, respectively. In the complete BEAST catalogs, we include a column called [\texttt{HRC}] to identify whether a source passes [\texttt{1}] or fails [\texttt{0}] these quality cuts.

In Figures \ref{fig:quality_cuts_cmd_smc} and \ref{fig:quality_cuts_cmd_lmc}, we show color-magnitude diagrams (CMD) in the UV (left), optical (middle), and IR (right) of sources in the SMC and LMC, respectively, that pass (top row) and fail (lower rows) each of the HRC cuts. Due to the magnitude overlap between these populations, we plot each population separately. In the upper row, we show the density distribution of sources that pass the HRC cuts compared to all the sources that failed (grey). Inversely, in the lower rows, we show the density distribution of sources that failed each specific HRC cut (F336W bias, F336W SNR, $A_V$ SNR), compared with sources that passed. We apply each cut independently to the original catalog, meaning the combined fractions of excluded sources should not add up to 100\%.

The F336W bias and F336W SNR cuts both remove a similar fraction of sources (52.6\% and 52.8\%, respectively) that are generally dimmer and span a broader range of colors. These cuts remove poorly modeled young stellar objects to the left of the lower main sequence \citep{Gouliermis2012}, and dim red background galaxies. In general, these two cuts remove sources below the average $50\%$ completeness limit of $m_{\rm F275W}=26.2$, $m_{\rm F336W}=26.9$, $m_{\rm F475W}=27.8$, $m_{\rm F814W}=25.5$, $m_{\rm F110W}=24.7$, and $m_{\rm F160W}=24.0$  (dotted lines).
Conversely, the $A_V$ SNR cut only removes 1.8\% of sources, however, these sources are generally much brighter and are likely either foreground stars blueward of the optical main sequence (as shown in Figure \ref{fig:eg_cmd}) or sources associated with bright diffraction spikes.

\begin{deluxetable*}{cc|ccccccc|cc} 
\tablecaption{\label{tab:filter_combos} Filter Combinations}
\tablehead{\colhead{Name} & \colhead{\# Filters} &  \colhead{F225W} & \colhead{F275W} & \colhead{F336W} & \colhead{F475W} & \colhead{F814W} & \colhead{F110W} & \colhead{F160W} & \colhead{$N_{\rm LMC}$} & \colhead{$N_{\rm SMC}$}}
\startdata
\textbf{A} & \textbf{7} & \xmark & \xmark & \xmark & \xmark & \xmark & \xmark & \xmark &  28&  4 \\ 
\textbf{B} & \textbf{6} &        & \xmark & \xmark & \xmark & \xmark & \xmark & \xmark &  2 &  6 \\ 
C & 6 & \xmark & \xmark & \xmark & \xmark & \xmark &        & \xmark &  2 &  2 \\ 
D & 5 &       & \xmark & \xmark & \xmark & \xmark &        & \xmark &  3 &  2 \\ 
S5& 5 &        & \xmark & \xmark & \xmark & \xmark & \xmark &        &  0 &  0 \\ 
\textbf{E} & \textbf{5} & \xmark & \xmark & \xmark & \xmark & \xmark &        &        &  2 &  4 \\ 
  & 4 &        & \xmark &        & \xmark & \xmark &        & \xmark &  1 &  0 \\ 
F & 4 &        &        & \xmark & \xmark & \xmark &        & \xmark &  2 &  3 \\ 
\textbf{G} & \textbf{4} &        & \xmark & \xmark & \xmark & \xmark &        &        &  7 &  4 \\ 
\textbf{H} & \textbf{3} &        &        & \xmark & \xmark & \xmark &        &        &  9 &  8 \\ 
\enddata  
\tablecomments{This table summarizes the unique filter combinations for the Scylla and METAL fields used in this work. The filter combination names are carried over from \citet{murray2024b}. The number of fields with each combination in the LMC ($N_{\rm LMC}$) and SMC ($N_{\rm SMC}$) is indicated in the right-hand columns. The bolded filter combinations are plotted in Figure \ref{fig:validation}.}   
\end{deluxetable*}

\begin{deluxetable*}{cc|ccccccc|c}[ht]
\tablecaption{BEAST Simulation Validation}\label{tab:val}
\tablehead{ 
 \multirow{2}{*}{Name} & \multirow{2}{*}{\# Filters} & $\Delta\log(M_{ini})$ & $\Delta\log(A)$ & $\Delta Z$ & 
 $\Delta$Distance & $\Delta A_V$ & $\Delta R_V$ & $\Delta f_\mathcal{A}$ & \multirow{2}{*}{$N_{stars}$} \\
 & & $(M_{\odot})$ & (years) & $(Z_{\odot})$ & (kpc) & (mag) & & &  }
\startdata
 \textbf{A} & \textbf{7} & 0.1821 & 1.3014 & 0.1326 & 15.6115 & 0.2285 & 0.4728 & 0.1714 & 8174 \\
 \textbf{B} & \textbf{6} & 0.1892 & 1.3099 & 0.1333 & 15.7091 & 0.2384 & 0.4808 & 0.1827 & 8168 \\
 C & 6 & 0.2013 & 1.3304 & 0.1461 & 15.7633 & 0.3366 & 0.4756 & 0.1760 & 8198 \\
 D & 5 & 0.2076 & 1.3388 & 0.1471 & 15.8588 & 0.3451 & 0.4828 & 0.1870 & 8187 \\
 S5 & 5 & 0.2127 & 1.3477 & 0.1489 & 15.9908 & 0.3196 & 0.5870 & 0.1984 & 8211 \\
 \textbf{E} & \textbf{5} & 0.2195 & 1.3582 & 0.1550 & 16.0989 & 0.3892 & 0.7586 & 0.2038 & 8231 \\
 F & 4 & 0.2349 & 1.3848 & 0.1742 & 16.1338 & 0.4982 & 0.5069 & 0.1968 & 8201 \\
 \textbf{G} & \textbf{4} & 0.2289 & 1.3749 & 0.1586 & 16.1850 & 0.4067 & 0.7917 & 0.2136 & 8232 \\
 \textbf{H} & \textbf{3} & 0.2864 & 1.4516 & 0.1830 & 16.3361 & 0.6458 & 0.9012 & 0.2245 & 8229 \\ 
\enddata
\centering
\tablecomments{Root mean square (RMS) errors for multi-band BEAST fits compared to inputs for $10^4$ simulated stars. Table \ref{tab:filter_combos} outlines filters used in each filter combination.}
\end{deluxetable*}

\begin{figure*}[t!]
 \centering
    \includegraphics[width=\textwidth]{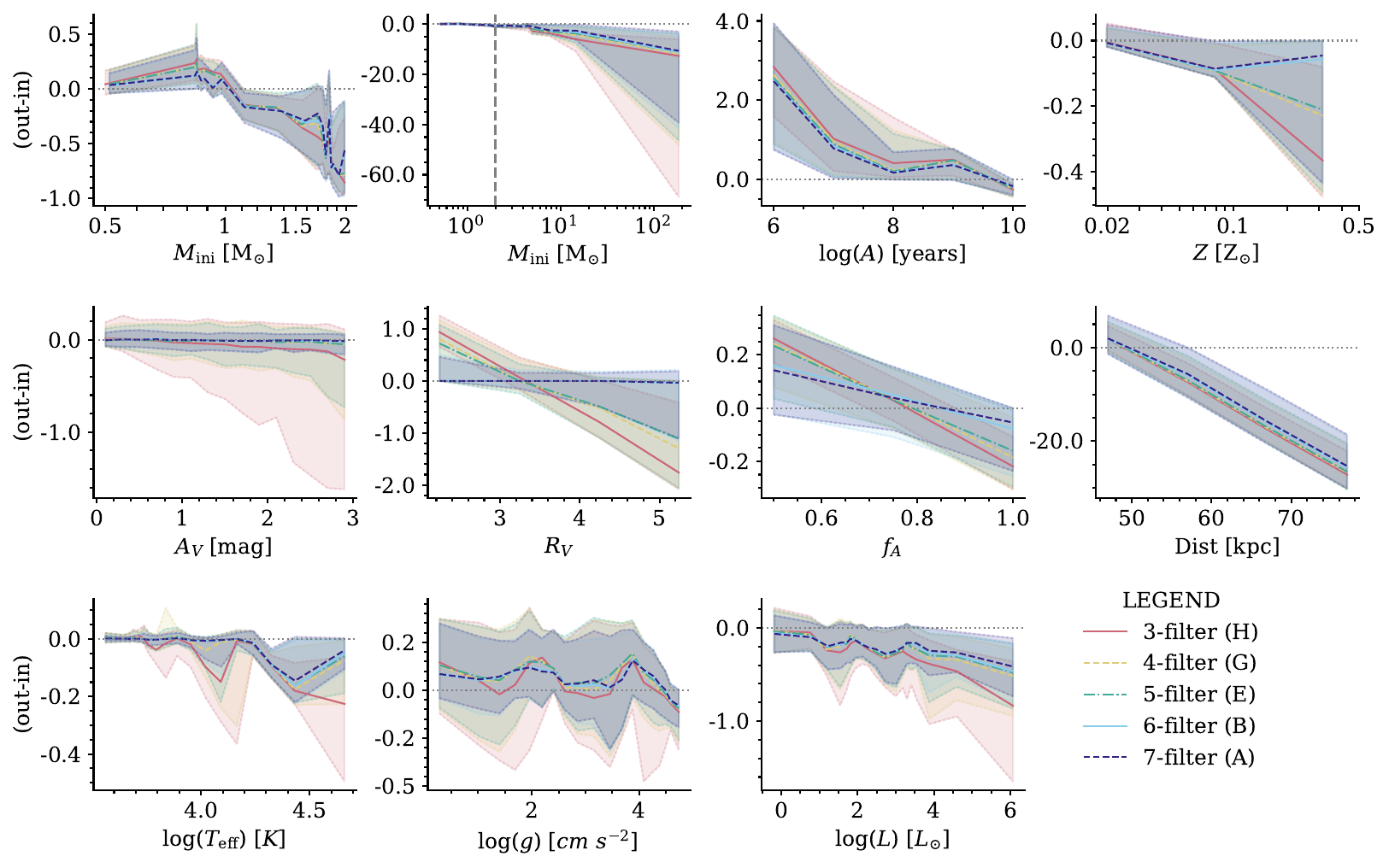}
	\caption{\textbf{Validation}: The median $\pm 1\sigma$ difference between the BEAST-modeled (``out'') and simulated (``in'') parameters versus the simulated parameters for $10^4$ stars simulated in 3, 4, 5, 6, and 7-filter cases, with high reliability cuts applied. For initial masses ($M_{ini}$), we show both the full input parameter range, and a zoom-in on sources with $M_{ini}\leq2M_{\odot}$ (dashed). Additional filters improve BEAST SED fits for nearly all parameters, especially $Z$, $A_V$, and $R_V$.}
	\label{fig:validation}
\end{figure*}

\section{Validation}\label{sec:validation}

Although a significant fraction of fields in this study were observed with 6-7 filters from the NUV-NIR, there are still many fields with only 3-4 filters of coverage in the UV and optical. To test the accuracy of the BEAST fitting method as a function of filter coverage, we construct simple simulations. First, we generate a grid of BEAST models using the same parameters as the BEAST model grid for the LMC (Table \ref{tab:beast}), with coarser resolution in all primary variables [$log(A)=1.0$ dex, $log(Z)=0.6\ Z_{\odot}$, $A_V=0.2$ mag, $R_V=1.0$, $f_\mathcal{A}=0.25$, $d=10$ kpc] and sample $10^4$ stars from this grid. Next, using the ASTs from field 14675{\textunderscore}LMC-15562nw-33974 (M$\_ 7$), which was observed in all seven filters, we generate an observation model to estimate photometric biases and uncertainties, and fit the simulated stars using the full-resolution BEAST model grid for the LMC. We repeat this procedure for nearly all combinations of Scylla filter coverage, summarized in Table \ref{tab:filter_combos}. After fitting, we apply the same high-reliability catalog cuts as described in Section \ref{sec:quality} to the simulated catalogs, leaving us with several thousand high-reliability sources for analysis in each case. 

Although the parameter range of the simulated stars is limited, these simulations give us an idea of what BEAST parameters we can reliably recover within the range of available Scylla and METAL filters. In Table \ref{tab:val}, we quantify the accuracy of each BEAST parameter for each filter combination by calculating the root mean square (RMS) error for all high-reliability source. We compute the RMS error for each BEAST parameter ($x$) as follows:
$$RMS(x) = \sqrt{\sum_{i=1}^{n_{\mathrm{HRC}}} \frac{(x_{i, out}-x_{i, in})^2}{n_{\mathrm{HRC}}}}$$ 
where $n_{\mathrm{HRC}}$ is the number of sources that pass the high-reliability cuts out of the original 10,000 stars simulated (Table \ref{tab:val}; right-most column). In instances where parameter values span a large dynamic range (e.g.~mass, age), we compute the RMS errors using the logarithmic parameter values. From these values, we see that measurements of certain parameters only improve marginally with additional filters (e.g.~age, distance, $f_\mathcal{A}$), whereas others improve significantly (e.g.~mass, $A_V$, $R_V$).

To visualize these trends, in Figure \ref{fig:validation}, we plot the differences between the input BEAST parameter values used to generate each simulated star and the output BEAST median values for five of the most common filter combinations, ranging from three to seven filters\footnote{Exact filters are detailed in Table \ref{tab:filter_combos}. We omit analyzing filter combinations that only occur in singular instances.}. We show both the stellar and dust parameters, and the derived stellar parameters ($T$, $g$, $L$).

With 3-filter coverage, we can reliably measure most stellar masses and older ages ($\log(A)>8$ years). Despite large uncertainties, we also reliably measure dust extinction ($A_V$), although there is systematic bias toward higher magnitudes, the potential source of which we will address later on in Section \ref{sec:binaries}. Most other BEAST parameters like metallicity ($Z$), $R_V$, $f_\mathcal{A}$, and distance, are poorly recovered with 3-filter coverage.

With additional filter coverage, we observe significant improvement in the measurement of certain parameters, such as metallicity ($N_{filters} \geq 6$), extinction ($N_{filters} \geq 4$), and $R_V$ ($N_{filters} \geq 6$). These filter requirements align with previous BEAST papers which required a minimum 4-band detection for stars in NGC 4214 \citep{choi2020}. Both $f_\mathcal{A}$ and distance remain poorly recovered, even with seven filters of observations.

Figure \ref{fig:validation} reveals systematic errors present in the BEAST output, notably overestimation at low input values (e.g.~age, $R_V$, $f_\mathcal{A}$, distance) and underestimation at high input values (e.g.~$M_{ini}$, $Z$, $R_V$, $f_\mathcal{A}$, distance). For a parameter like initial mass with an IMF-based prior, the default expectation is that most observed sources will be lower mass. Therefore, if a more massive star is poorly-constrained, its mass will be underestimated. Cumulatively, this results in a systemic underestimation, which we see for BEAST initial mass values greater than one solar-mass. The same principle applies to BEAST age estimates, which assumes a flat star-formation history prior, resulting in an overestimation of ages for young stars.
At high metallicities, underestimation is likely due to edge effects in the model gridding. For example, stars simulated with $Z=0.5\ Z_{\odot}$ will intrinsically be fit with some level of scatter, both greater than and less than $0.5\ Z_{\odot}$. However, in the absence of any models with $Z>0.5\ Z_{\odot}$, only the lower scatter will be captured in the distributions, resulting in a systematic underestimation at grid edges. This edge effect is also present in the $R_V$, $f_\mathcal{A}$, and distance results, where lower values are systematically overestimated and high values are systematically underestimated.

In all filter scenarios, distances remain poorly constrained, with distances systematically trending towards 50 kpc regardless of input values, despite providing a flat distance prior. At this time, it is unclear what the source of this systematic trend is. These fits could potentially be improved by providing a more finely spaced distance grid, or by iteratively updating distance fits by collectively modeling the distance distribution of all sources in a field (which will be possible with the MegaBEAST\footnote{https://github.com/BEAST-Fitting/megabeast}).

Due to the variable quality of parameters based on the number of filter coverage, some discussions will focus on parameter-optimized subsets of the observed fields. For example, $Z$ and $R_V$ measurements are only well constrained in the 6-7 filter cases, so when discussing $Z$ and $R_V$ results in detail, we will only use measurements from fields with 6 or 7 filters. Similarly, discussions of extinction results will emphasize data from fields observed with at least 4 filters to avoid systematic bias from 3-filter fits. 

\begin{figure*}[h]
    \centering
    \includegraphics[width=1.03\textwidth]{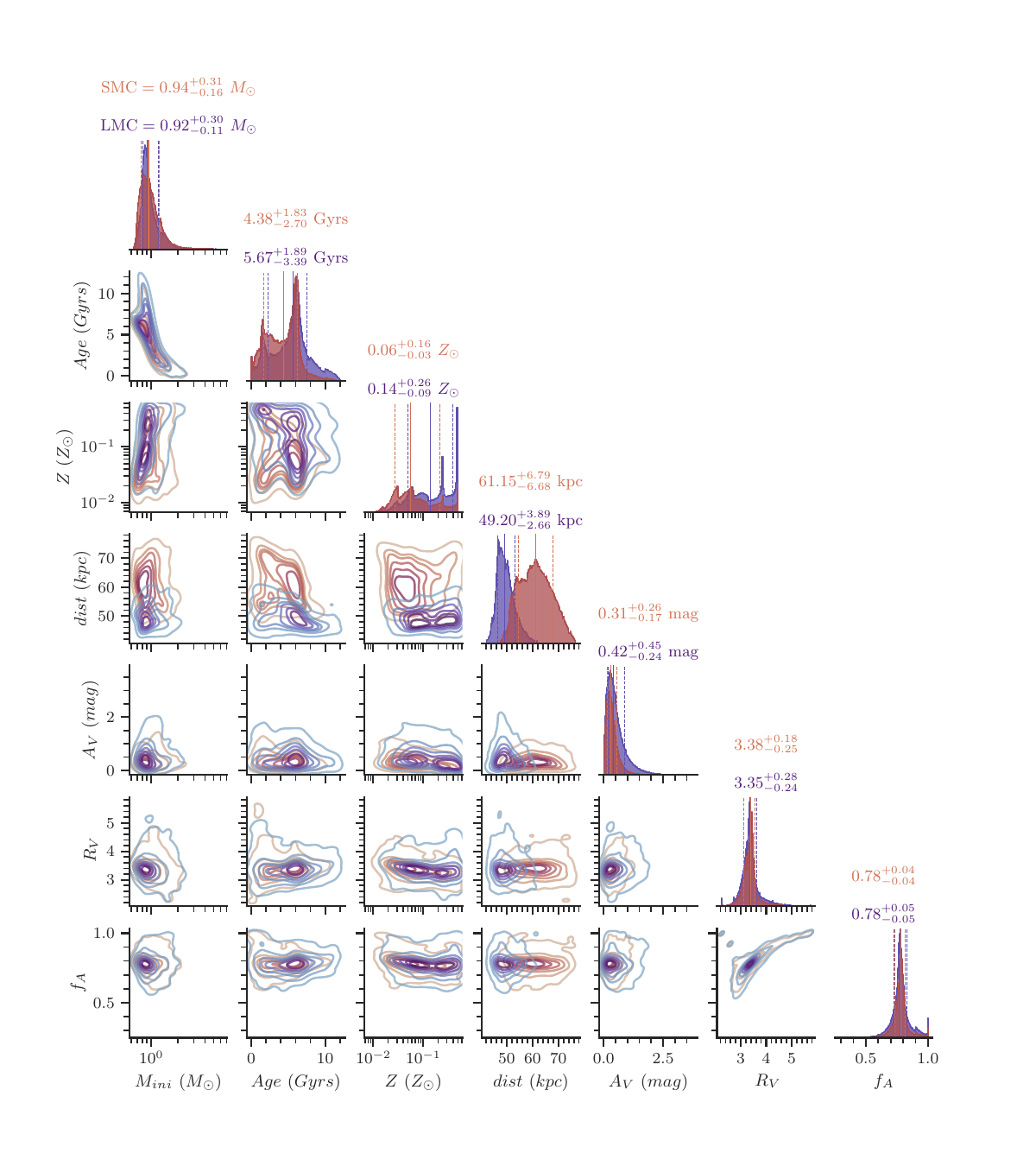}
    \caption{\textbf{BEAST parameter results for all HRC sources in the LMC and SMC}: 2D distributions of the median \texttt{p50} values for all BEAST primary stellar parameters - initial mass ($M_{ini}$), age ($A$), metallicity ($Z$), distance - and dust parameters  - $A_V$, $R_V$, $f_\mathcal{A}$, for all high-reliability sources in the LMC (blue) and SMC (red). 1D histograms of all parameters are shown along the diagonal along with the corresponding median and $1\sigma$ values for the SMC and LMC, respectively.}
    \label{fig:beast}
\end{figure*}

\section{Results}
\label{sec:results}

In this section, we present the BEAST fits in detail, describe any notable systematics, and discuss the differences between fits in the two galaxies. The BEAST catalogs are made publicly available as High-Level Science Products (HLSP) on MAST\footnote{https://archive.stsci.edu/hlsp/scylla}. For all parameter values, we use the reported median values (\texttt{p50}), as opposed to the singular best-fit model values (\texttt{Best}) or the expectation values  (\texttt{Exp}). We opt to use the \texttt{p50} values since these measurements are calculated using the entire posterior grid, meaning we can also estimate parameter uncertainties from the width of the marginalized 1D pPDF.

In Figure \ref{fig:beast}, we plot the 2D distributions of the median \texttt{p50} values for all primary stellar (initial mass, age, metallicity, distance) and dust ($A_V$, $R_V$, $f_\mathcal{A}$) parameters for all high-reliability sources in the LMC (blue) and SMC (red). Correlations reflect trends in the stellar population. Along the diagonal, we plot the 1D histograms of all the parameters individually, along with the corresponding median and $1\sigma$ values for the two galaxies. 

\paragraph{Initial Mass and Age} Similar stellar initial mass ($M_{ini}$) and age ($logA$) distributions are observed in the stellar populations of the two galaxies, with median and standard deviation initial mass estimates of $0.94^{+0.31}_{-0.16}$ ($0.92^{+0.30}_{-0.11}$) $M_\odot$ and ages of $4.38^{+1.83}_{-2.70}$ ($5.67^{+1.89}_{-3.39}$) Gyrs in the SMC (LMC), respectively. There is a strong correlation between initial mass and age which is expected since higher-mass stars live shorter life spans than low-mass stars. We anticipate few BEAST fits of massive stars since these stars are often saturated in the photometry (see Figure \ref{fig:eg_cmd} in Section \ref{sec:data} for further details). 

\paragraph{Metallicity} Although the metallicity grid was the same for both galaxies, stars in the LMC are on average fit to be more metal-rich than the SMC, with median metallicity fits of $14^{+26}_{-9}$\% $Z_{\odot}$  ([M/H] = $-0.71^{+ -0.44}_{--0.90}$, assuming $\alpha/Fe=0.2$) in the LMC versus $6^{+16}_{-3}$\% $Z_{\odot}$ ([M/H] = $-1.07^{+-0.65}_{--1.37}$) in the SMC. The 1D metallicity histogram for the LMC features several spikes, particularly at $20$\% and $50$\% $Z_{\odot}$. The metallicity grid was originally constructed to span from $\sim1$\% solar to $\sim50$\% $Z_{\odot}$, based on the prior knowledge of the metallicities in the MCs. However, the pile-up of LMC sources at $\sim50$\% $Z_{\odot}$ indicates that fits would likely have improved had higher metallicities been included. Additionally, simulations from Section \ref{sec:validation} show that metallicity is systematically underestimated when fitting with only optical and UV filters (i.e.~$3\mbox{--}5$ filters), meaning many of the sources in Figure \ref{fig:beast} would likely have been fit as higher metallicity had additional IR filters been available. Future survey fittings and releases will account for this issue and expand the metallicity range. 

\paragraph{Distance} The distance grid was intentionally different for the two galaxies, with the LMC grid ranging from $40-60$ kpc, and the SMC grid ranging from $47-77$ kpc. This was done to account for the prior distance estimates for the galaxies, with the LMC at $49.89$ kpc and the SMC at $61.94$ kpc, respectively \citep[][]{degrijs2014, degrijs2015}. We expect a broad distance distribution for both galaxies, especially for the SMC which is quite extended along the line-of-sight \citep[reviewed in][]{murray2024a}.

For the LMC, BEAST distance estimates span the full grid with a median distance fit of $49.2^{+3.9}_{-2.7}$ kpc, however, the distribution of distances is slightly asymmetric, skewed towards larger distances. This distance distribution agrees with depth measurements from \citet{Subramanian2009} which found an average depth of $4.0\pm1.4$ kpc in the bar region and $3.4\pm1.2$ kpc in the disk. We believe the BEAST-derived distance distribution might be slightly smaller since we measure both young and old stars, whereas \citet{Subramanian2009} only measured the distances to red clump stars, which are generally older and would therefore have a larger scale height.

In the SMC, distance estimates also span the full grid, with a median distance fit of $61.2^{+6.8}_{-6.7}$ kpc, however, the distance distribution is not symmetric and there is evidence of bimodality in the distribution. These results corroborate other studies that show the SMC comprises two star-forming systems separated by several kpc along the line-of-sight \citep{Subramanian2017, murray2024a}. This multi-components structure could potentially be the result of tidal stripping due to a recent near impact ($\sim5$ kpc) with the LMC roughly 150 Myrs ago \citep{choi2022}.

While median BEAST distance fits do agree with existing distance measurements, validation tests (Section \ref{sec:validation}) showed that individual distance fits are often poorly recovered, spanning $\pm5\mbox{--}10$ kpc in uncertainty. One potential explanation for this paradox is that individual stars might not contain enough information to recover distances, however, as an ensemble with thousands of sources, we can recover signal, even if individual sources have low signal-to-noise. As such, we caution anyone wanting to use distance measurements for individual stars to be careful in how they choose to interpret these measurements.

\paragraph{Extinction} Both galaxies exhibit low amounts of extinction ($A_V$) along the line-of-sight towards stars, with median extinction values of $\sim0.31^{+0.26}_{-0.17}$ ($0.42^{+0.45}_{-0.24}$) mag in the SMC (LMC), respectively. These extinction measurements include contributions for Milky Way foreground dust. The extinction distributions in both galaxies follow log-normal distributions, with tails leading towards higher extinctions. The log-normal distribution in the LMC is broader than in the SMC.

\paragraph{Dust Properties} The effects of extinction are more influential on the observed SEDs than other dust parameters like $R_V$ and $f_\mathcal{A}$. As a result, the influence of these secondary dust parameters is far harder to constrain without high-quality observations farther into the UV (e.g.~F275W and F225W). When observations lack enough information to inform the likelihood, the posterior distribution will default to the prior distribution. The BEAST model grid adopted the same 2D prior for the secondary dust parameters $R_V$ and $f_\mathcal{A}$ as described in Figure 7 in \citet{gordon2016}, with a peak in the marginalized $R_V$ distribution around $\sim2.7$ and $f_\mathcal{A}$ at 1.0. Given the resolution of the parameter gridding, the weighted average of this 2D prior falls near [3.55, 0.73] which is near what most of the observations indicate in both galaxies with median $R_V$ fits of $3.38^{+0.18}_{-0.25}$ ($3.35^{+0.28}_{-0.24}$) and median $f_\mathcal{A}$ fits of $0.78^{+0.04}_{-0.04}$ ($0.78^{+0.05}_{-0.05}$) in the SMC (LMC). Lack of variation across both galaxies and the similarity to the 2D prior indicates most $R_V$ and $f_\mathcal{A}$ measurements are poorly-constrained in most fields. Individual fields with coverage in multiple UV filters (F275W, and F336W) could potentially constrain variations in $R_V$, something future Scylla science publications will investigate (Murray et al. 2025, in prep.).

\begin{figure*}[!ht]
    \centering
    \includegraphics[width=0.9\textwidth]{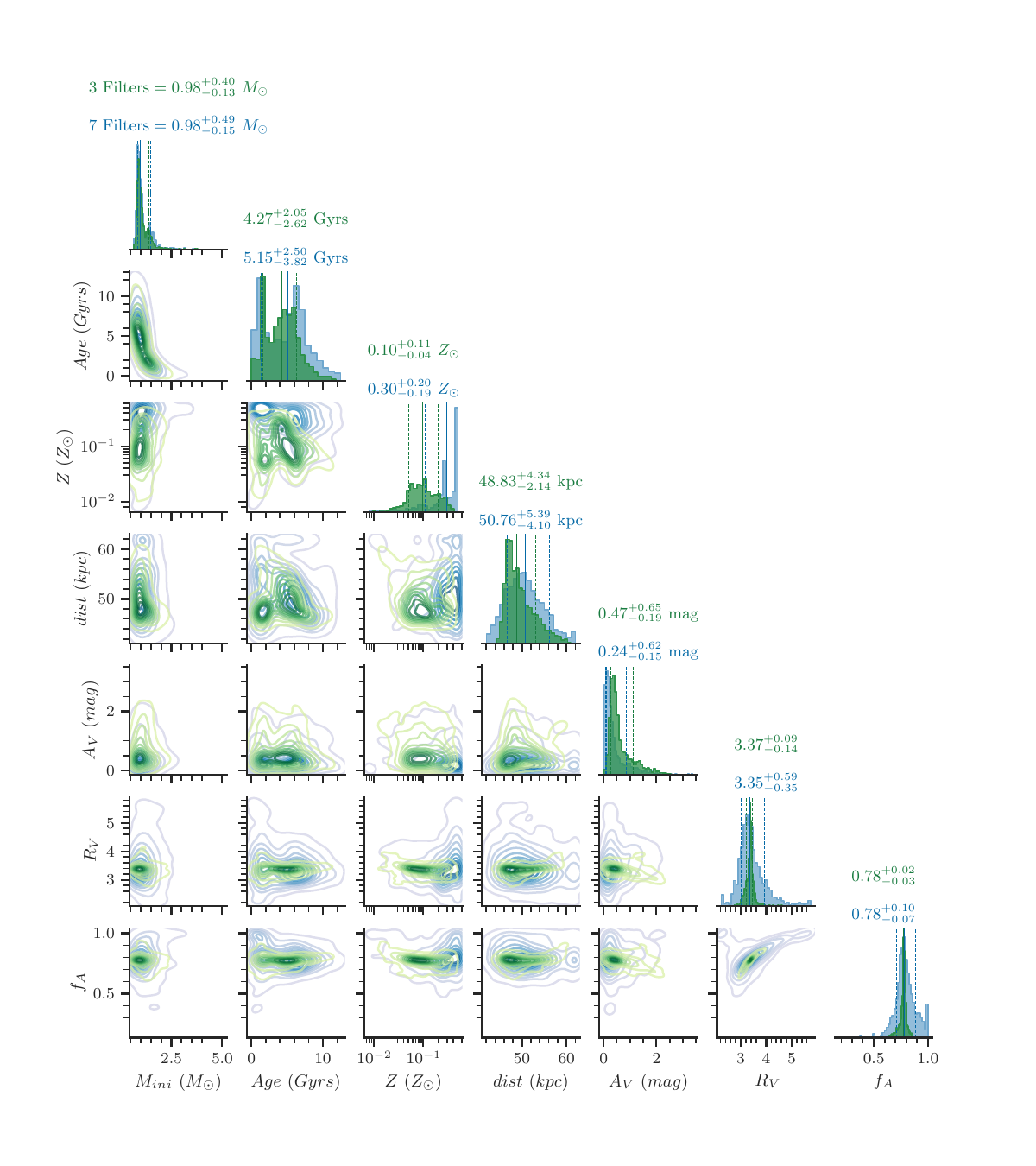}
    \caption{\textbf{Filter impact on BEAST parameter results (3 vs.~7)}: 2D distribution of BEAST parameters for an example field in the LMC (14675$\textunderscore$LMC-15562nw-33974; M6). The field was originally observed and fit with 7 filters (blue) and BEAST fits were repeated using 3 filters (F336W, F475W, F814W; green) for comparison. Some BEAST parameters, like initial mass ($M_{ini}$) and age, are measured to have similar marginalized distributions with both 3- and 7-filter coverage. Other parameters like metallicity ($Z$), however, differ significantly depending on the number of filters. In particular, the median extinction ($A_V$) measured with 3-filter coverage is nearly double the median measured with 7-filter coverage. }
    \label{fig:metal6a}
\end{figure*}

\begin{figure*}[!ht]
    \centering
    \includegraphics[width=0.9\textwidth]{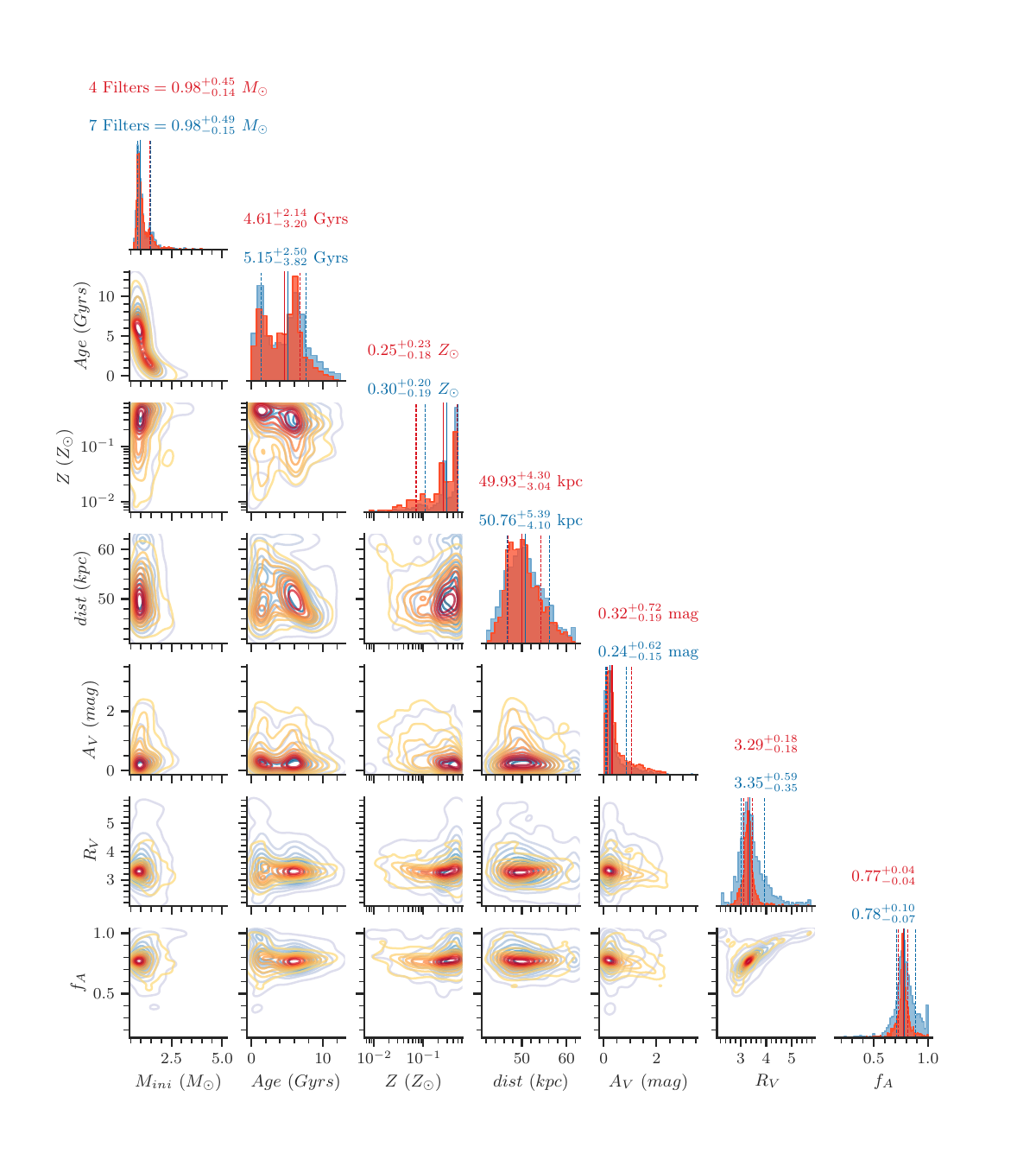}
    \caption{\textbf{Filter impact on BEAST parameter results (4 vs.~7)}: 2D distribution of BEAST parameters for an example field in the LMC (14675$\textunderscore$LMC-15562nw-33974; M6). The field was originally observed and fit with 7 filters (green) and BEAST fits were repeated using 4 filters (F275W, F336W, F475W, F814W; red) for comparison. The inclusion of F275W increases BEAST metallicity measurements significantly compared to 3-filter fits (Fig.~\ref{fig:metal6a}).}
    \label{fig:metal6b}
\end{figure*}


\begin{figure*}[!ht]
    \centering
    \includegraphics[width=\textwidth]{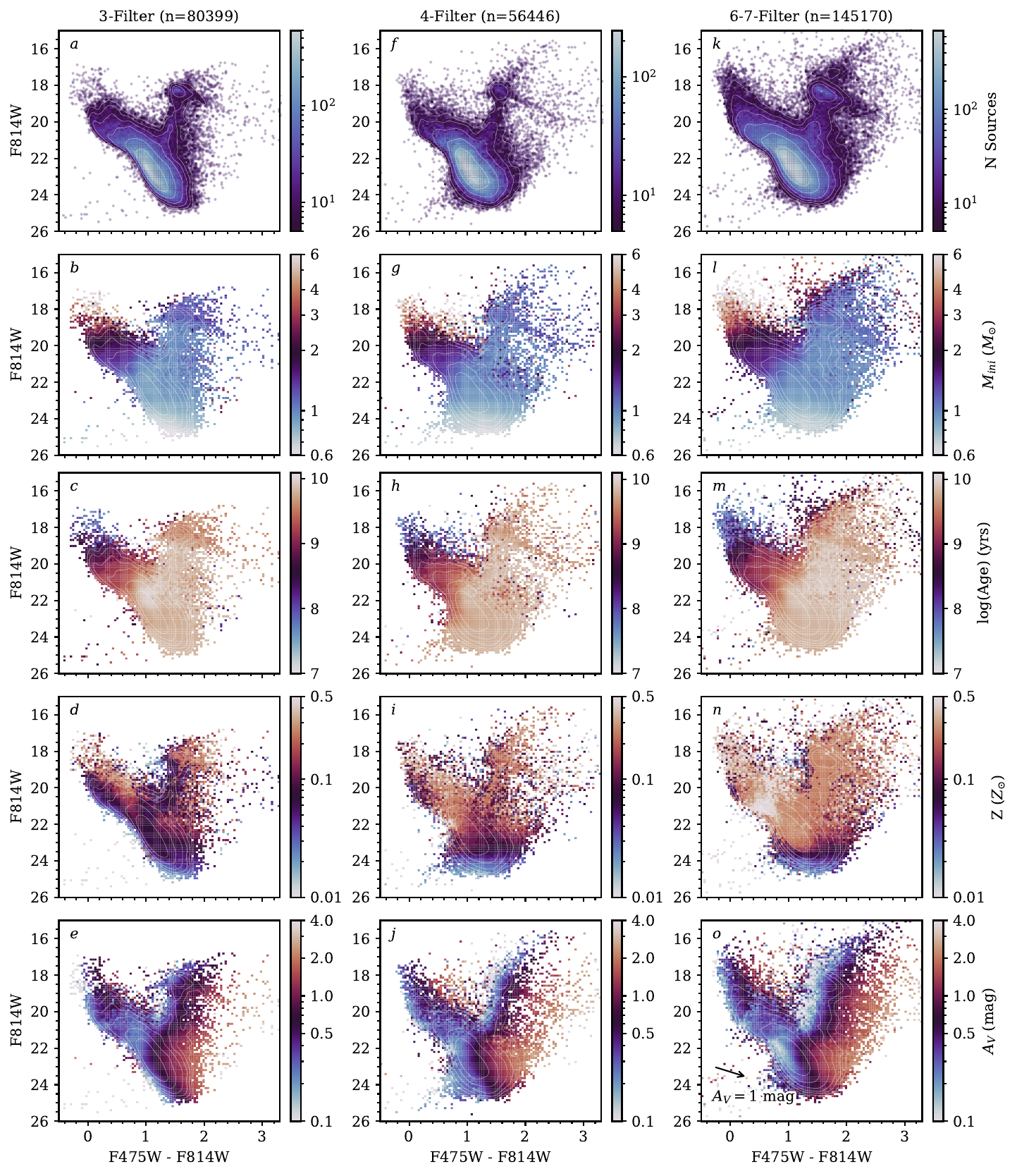}
    \caption{\textbf{CMD BEAST trends in the LMC}: Optical CMDs for all high-reliability sources in the LMC from fields with 3-filter coverage (left; $n=80,399$), 4-filter coverage (middle; $n=56,446$), and 6-7-filter coverage (right; $n=145,170$). Top columns ($a, f, k$): Density distribution of sources as colored 2D histograms, and outlined in logarithmic levels using kernel-density estimates (KDE). KDE density outlines (white) are propagated to all other rows. All other rows: Spatially binned ($100\times100$) medians for notable BEAST parameters: initial mass ($M_{ini}$), age, metallicity ($Z$), and extinction ($A_V$). In panel $o$, we show the direction of a 1-magnitude reddening vector with a black arrow. UV and IR versions of the 6-7-filter sources are provided in Appendix \ref{apx:cmd}.}
    \label{fig:beast_cmd_lmc}
\end{figure*}

\begin{figure*}[!ht]
    \centering
    \includegraphics[width=\textwidth]{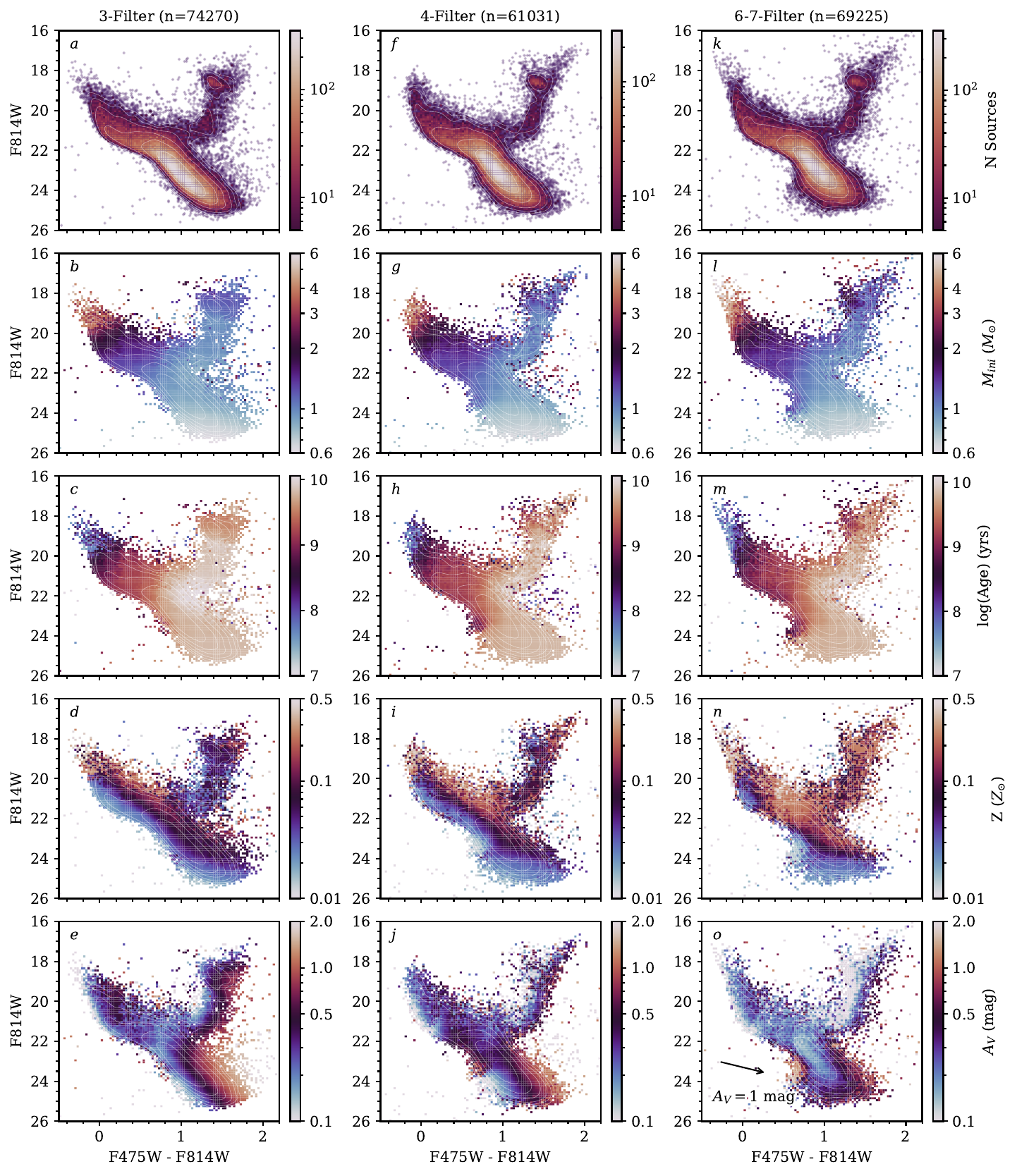}
    \caption{\textbf{CMD BEAST trends in the SMC (Same as Fig.~\ref{fig:beast_cmd_lmc})}: Optical CMDs for all high-reliability sources in the SMC from fields with 3-filter coverage (left; $n=74,270$), 4-filter coverage (middle; $n=61,031$), and 6-7-filter coverage (right; $n=69,225$). Notable differences from Fig.~\ref{fig:beast_cmd_lmc}: extinction color range has been reduced to improve dynamic range.}
    \label{fig:beast_cmd_smc}
\end{figure*}

\subsection{Filter Variation}

There are considerable variations in the BEAST results and uncertainties when fitting sources with 3 versus 4 versus 7 bands of observations. In this section, we investigate these variations on an individual Scylla field and comment on their application to the broader results of the paper. We use 14675$\textunderscore$LMC-15562nw-33974 (M6) to investigate the effects of filter variation on BEAST results. This field is located in the upper west corner of the LMC and was originally observed in seven filters. Although this field spatially coincided with two other Scylla fields, LMC$\textunderscore$8 and M5, with three- and six-filter coverage, respectively, we find that minor differences in the photometry between each field make it difficult to isolate the effects of filter coverage. For a true apples-to-apples comparison, we instead chose to refit the M6 field using data from only 3-and 4-filter combinations (H and G in Table \ref{tab:filter_combos}). We apply HRC cuts to all BEAST catalogs, retaining 1646, 1670, and 1614 sources in the 3-, 4-, and 7- filter catalogs, respectively, out of the total 5342 sources with BEAST fits.

In Figures \ref{fig:metal6a} and \ref{fig:metal6b}, we plot the 2D distribution of BEAST parameters for all HRC sources in the 3-filter catalog (green), 4-filter catalog (red), and 7-filter catalog (blue). By overlaying the 1D and 2D distributions, we can see that the distributions of certain BEAST parameters vary considerably depending on the number of filters. Most notably, stellar metallicity is measured to be much lower when fit with only three filters, compared to seven filters. Meanwhile, the median extinction tends to be double when measured with three filters compared to seven. From these trends, we surmise that any 3-filter fields with low metallicity estimates are likely incorrect and that median galactic estimates of extinction in Figure \ref{fig:beast} are likely overpredicted due to contributions from 3-filter fields. From Figure \ref{fig:metal6b}, we see that the inclusion of even one additional filter (F275W) increases metallicity measurements for nearly all sources. In addition, median extinction measurements are lower, however there is still a slight overestimation compared to 7-filter measurements. 



\section{Discussion}
\label{sec:discussion}

With high-reliability stellar and dust fits for over 550,000 sources, we can now analyze what BEAST results tell us about the stellar and dust content of the clouds and how it compares to existing literature. In this section, we first discuss the stellar content of these Scylla fields, see how BEAST-fit mass estimates compare with isochrone-predicted masses, and discuss the potential influence of binary populations (Section \ref{sec:content}). Second, we compare the BEAST age distributions with existing star-formation history measurements (Section \ref{sec:sfh}). Third, we quantify our sensitivity to $A_V$ as a function of stellar luminosity, characterize the log-normal extinction distributions in each galaxy, and compare the average extinction values with other ISM tracers (Section \ref{sec:extinction}). 

\subsection{Stellar Content}\label{sec:content}

A galaxy consists of many types of stellar populations which can be used to address broader science questions about the galaxy itself: e.g.~initial mass function \citep{weisz2013}, star-formation history \citep{lewis2015}, disk structure \citep{Williams2012}, metallicity distribution \citep{Gregersen2015}, total column extinction \citep{dalcanton2015} - as well as more targeted questions about physical processes in galaxy evolution: e.g.~AGB stars as a source of dust production \citep{Goldman2022} or formation of stellar clusters \citep{johnson2017}. 

From CMDs alone, we can identify several stellar populations based on their locations within color-magnitude space. In Figures \ref{fig:beast_cmd_lmc} and \ref{fig:beast_cmd_smc}, we plot optical CMDs for all high-reliability sources in the LMC and SMC, respectively. The sources in the CMDs are split up by the number of filters to help compare how variable filter coverage influences the measured BEAST parameters. The filter numbers were chosen such that similar numbers of sources were visualized in each CMD.

In the first row of Figures \ref{fig:beast_cmd_lmc} and \ref{fig:beast_cmd_smc} ($a,f, k$), we show the density distribution of all sources n fields with 3-, 4-, and 6-7-filter coverage in each galaxy, both as colored 2D histograms and outlined in logarithmic levels using kernel-density estimates (KDE). These KDE outlines are propagated to all other rows to provide a reference for where the majority of stars are located. Both galaxies feature a prominent main sequence, running from the top left (F475W-F814W$ = -0.5$, F814W$=18$ mag) down to the lower right (F475W-F814W $= 1.5/1.8$, F814W$=25$ mag). Perpendicular to this structure is the red giant branch (RGB), which extends upwards from F814W$=23$ mag to 17 mag. Along this branch is the red clump (RC) at $\sim$19 mag, which can be seen as an over-density along the RGB both in the SMC and LMC. 

While the morphology of the RGB and RC can be intrinsically complex \citep{girardi2016}, their overall distribution on a CMD contains valuable information about a galaxy, especially extinction which can cause considerable broadening of these CMD features. This is especially notable in the LMC, where the RC structure appears to be elongated down and to the right. Based on extinction coefficients for these optical bands, we can predict the reddening effect of some magnitude of extinction, also referred to as a reddening vector. This reddening vector for $A_V = 1$ mag is shown in the bottom-right column. The width of the LMC RC, compared to the orientation of the reddening vector shows that the elongation of the LMC RC is almost certainly the result of dust extinction.

In all subsequent rows of Figures \ref{fig:beast_cmd_lmc} and \ref{fig:beast_cmd_smc} (subfigures $b$-$e$, $g$-$j$, and $l$-$o$), we plot the spatially binned medians for primary BEAST parameters.

\paragraph{Initial Mass} In Figures \ref{fig:beast_cmd_lmc} and \ref{fig:beast_cmd_smc} ($b, g, l$), we plot the average initial mass estimates. The smallest/largest initial mass estimates for sources in the SMC and LMC are $0.58^{+0.05}_{-0.04}$/$24.39^{+1.60}_{-6.74} M_{\odot}$ and $0.55^{+0.03}_{-0.02}$/$51.27^{+14.29}_{-2.99} M_{\odot}$ respectively, however, for visual purposes, we limit the color range to $0.6\mbox{--}6\ M_{\odot}$. The main sequence spans a wide range of initial masses, whereas the RGB and RC are primarily composed of sources with initial masses $\sim1\ M_{\odot}$, consistent with expectations \citep[e.g.~][]{girardi2016, nataf2021}. Fields with 6-7-filter coverage contain more high-mass stars than fields with 3-4 filter coverage. This is likely the result of multiple variable like source saturation without guard exposures and spatial location of 6-7-filter fields near star-forming regions.

\paragraph{Stellar Age} In Figures \ref{fig:beast_cmd_lmc} and \ref{fig:beast_cmd_smc} ($c, h, m$), we plot the average ages of stars across the CMD. We find that age is spatially correlated with initial mass, i.e. higher-mass stars ($M> 4\ M_{\odot}$) tend to be less than 100 Myrs old, whereas lower-mass ($M < 1\ M_{\odot}$) tend to be greater than 10 Gyrs old. For higher-mass stars in the SMC, there is a gradient in age, with redder stars being more evolved. 

\paragraph{Metallicity} In Figures \ref{fig:beast_cmd_lmc} and \ref{fig:beast_cmd_smc} ($d, i, n$), we plot the average metallicities and find notable differences between the distributions depending on the number of filters in both the SMC and the LMC. In the SMC, there is a gradient in metallicity from $3\mbox{--}20\%\ Z_{\odot}$, perpendicular to the main sequence where redder sources are more metal-rich. This metallicity gradient could explain why the single-metallicity isochrones plotted in Figure \ref{fig:eg_cmd} did not perfectly match the observed population distribution on the CMD. By comparison, the median metallicity for most sources in the RGB and RC is $\sim10\%\ Z_{\odot}$, similar to spectroscopic studies from APOGEE of RGB stars in the LMC \citep{povick2024}. In the LMC, the median metallicity for main sequence stars is more homogeneous ($20\mbox{--}30\%\ Z_{\odot}$), and both MS and RGB/RC metallicities are on average much greater than the SMC, $25\%\ Z_{\odot}$ compared to $10\%\ Z_{\odot}$. In both galaxies, high-mass stars ($M > 4\ M_{\odot}$) show the highest amount of enrichment, at $50\%\ Z_{\odot}$, whereas low-mass stars ($M<0.7\ M_{\odot}$) are the least enriched ($<3\%\ Z_{\odot}$). 

\paragraph{Extinction} In Figures \ref{fig:beast_cmd_lmc} and \ref{fig:beast_cmd_smc} ($e,j,o$), we plot the average extinction and show the direction of a 1-magnitude reddening vector. The LMC contains far greater amounts of extinction ($0.3\mbox{--}3.0$ mag) than the SMC ($0.2\mbox{--}1.0$ mag). The elongated RC and RGB in both galaxies show a gradient in extinction, ranging from $0.1\mbox{--}0.7$ mag in the SMC, which is in agreement with existing RGB $A_V$ estimates in the SMC of $\langle A_V\rangle = 0.41 \pm 0.09$ mag \citep{ymj2021}, and $0.1\mbox{--}4.0$ mag in the LMC. Surprisingly, the 3-filter SMC contains lower main sequence sources with extinctions greater than the maximum extinction experienced by the RGB: $A_{V} = 1.7$ mag for the lower main sequence versus $A_{V} = 0.7$ mag for the RGB. In Section \ref{sec:binaries}, we discuss the credibility of these $A_V$ estimates and show how binary stars could influence these results.

\begin{figure}
    \centering
    \includegraphics[width=\linewidth]{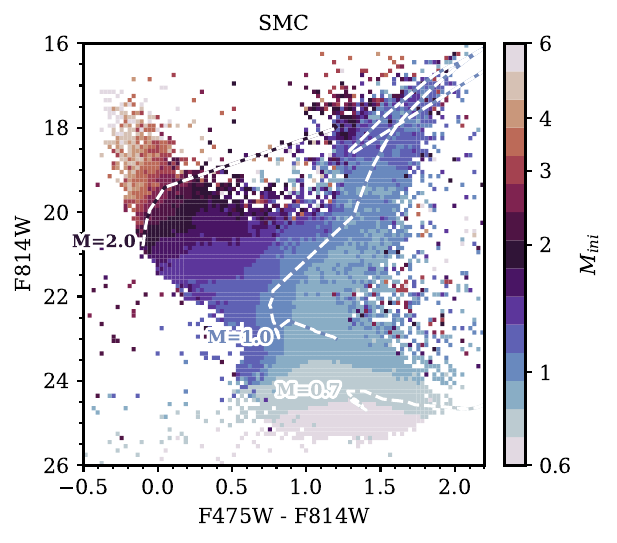}
    \caption{\textbf{Lower-mass limit}: Spatially binned optical CMD of all high-reliability sources in the SMC, colored by median initial mass. PARSEC stellar evolution tracks for stars with initial masses of 0.7, 1.0, and 2.0 $M_{\odot}$ are overlaid, assuming no extinction and $Z=20\%\ Z_{\odot}$ \citep{bressan2012}. We compare stellar evolutionary tracks with the SMC because it has a smaller range of metallicities and less extinction than the LMC. BEAST initial mass estimates are in broad agreement with evolutionary tracks.}
    \label{fig:iso}
\end{figure}

\subsubsection{Evolutionary Track Mass Estimates}\label{sec:mass}

Due to the deep high-resolution photometry from \textit{HST}, we resolve sub-solar mass sources. To validate mass estimates from the BEAST, we compare the CMD distribution of BEAST initial mass estimates in the SMC with PARSEC stellar evolutionary tracks \citep{bressan2012}. We opt to only make this comparison in the SMC since the LMC has much more extinction and a larger range of metallicities which complicates this comparison.

In Figure \ref{fig:iso}, we plot the spatially binned optical CMD of all high-reliability sources in the SMC, colored by the median initial mass estimates from the BEAST. This figure is the same as Figure \ref{fig:beast_cmd_smc}b, g, and l, combined, except the color bar now shows discrete mass bins to aid in the comparison. PARSEC stellar evolutionary tracks are overlaid, with age ranging from 10 Myrs (log(Age)$=7$ yrs) to 6.3 Gyrs (log(Age)$=9.8$ yrs) for stars with three initial masses: 0.7, 1.0, and 2.0 $M_{\odot}$. These models assume a metallicity of 20\% $Z_{\odot}$, a distance of 60 kpc, and apply no extinction. 

The evolutionary tracks for stars with initial masses of 0.7 and 1.0 $M_{\odot}$ are in agreement with the BEAST-predicted initial mass estimates. The 2.0 $M_{\odot}$-track also overlaps with BEAST estimates, however, the track seems to overlap more with slightly higher-mass stars ($M=2.5\mbox{--}3.0\ M_{\odot}$), and there is less agreement given the limited spatial coverage of the CMD. There are two potential reasons for the lack of alignment between higher-mass stars and their evolutionary tracks: (1) it could be that higher-mass stars are located in dusty environments, which would make the observed stars dimmer and redder than the stellar evolutionary track; (2) it could be that the higher-mass stars have metallicities greater than 20\%, which would make them dimmer and redder for a given mass on the main sequence. From Figure \ref{fig:beast_cmd_smc}, we know main sequence stars with 2.0 $M_{\odot}$ in the SMC exhibit a gradient of metallicities, ranging from $3\mbox{--}30\%\ Z_{\odot}$, while extinction varies from $0.1\mbox{--}0.5$ mag, so both of the effects noted above are likely responsible for the offset between the observed stars and the stellar evolutionary tracks at higher masses. Overall, the BEAST initial mass estimates agree with stellar evolutionary tracks.

\begin{figure}
    \centering
    \includegraphics[width=\linewidth]{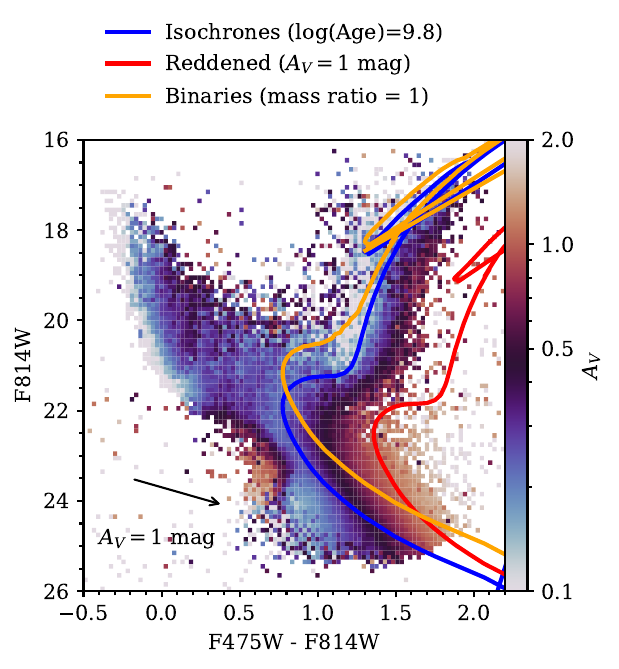}
    \caption{\textbf{Effects of binaries}: Spatially binned optical CMD of all high-reliability sources in the SMC, colored by median BEAST extinction. In the lower left, we show the direction of a 1-magnitude reddening vector as a black arrow. PARSEC isochrones for log(Age)$=9.8$ yrs with $Z=20\%\ Z_{\odot}$ and no extinction are overlaid in blue \citep{bressan2012}. Applying $A_V = 1$ mag produces the isochrone shown in red. Assuming all stars are binaries with an equal mass companion (binary mass ratio $= 1$) produces the isochrone shown in orange. The BEAST does not currently model include binary models, however, both effects are capable of populating the redder lower main sequence.}
    \label{fig:binaries}
\end{figure}

\begin{figure*}
    \centering
    \includegraphics[width=\textwidth]{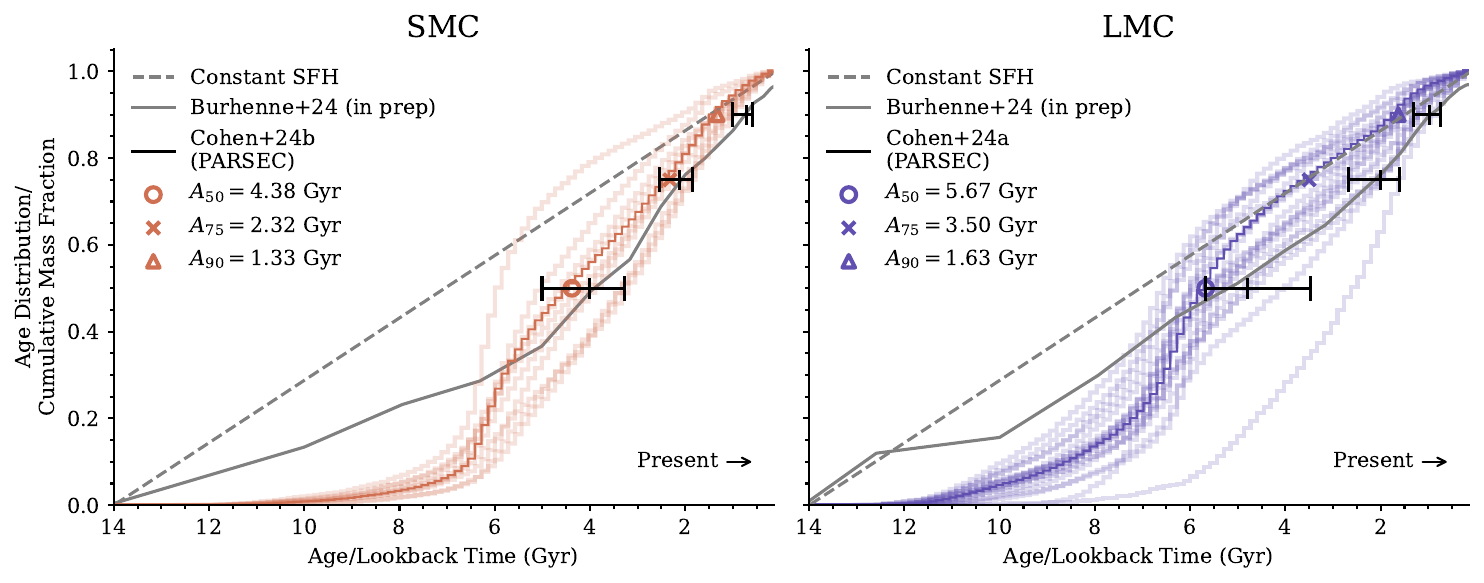}
    \caption{\textbf{Stellar age versus star-formation history}: Cumulative stellar age distributions of sources in each Scylla field in the SMC (left; red) and LMC (right; blue). The 50th (\texttt{o}), 75th (\texttt{x}), and 90th (\texttt{$\triangle$}) percentiles of the cumulative stellar age distributions for all HRC sources in the LMC and SMC are marked. We overlay median $\pm 1\sigma$ $\tau_{50}$, $\tau_{75}$, and $\tau_{90}$ measurements for all Scylla and METAL field from \citet{cohen2024a, cohen2024b} (black) and the total cumulative star formation histories for the SMC and LMC from (Burhenne et al. 2025, in prep.) (grey). For comparison, the distribution of a constant star-formation rate is shown as a dashed line. Overall, the distributions of age estimates from the BEAST are in broad agreement with independent SFH estimates, especially for the SMC.}
    \label{fig:sfh}
\end{figure*}

\subsubsection{Binary Stars}\label{sec:binaries}

As noted in earlier discussions of extinctions, parts of the lower main sequence in Figure \ref{fig:beast_cmd_smc}e seemed to experience double as much extinction as the RGB: $A_{V, max} = 1.7$ mag for the lower main sequence versus $A_{V, max} = 0.7$ mag for the RGB. Given their bright luminosities and extended scale heights relative to most star-forming galaxy disks, RGB sources are often used to quantify the total column extinction of external galaxies \citep[e.g.][]{dalcanton2015, ymj2021}. The $A_V$ range of the RGB in the SMC should capture the average range of extinctions within the galaxy. It is therefore unexpected to observe dim lower main-sequence stars with more than double the extinction, according to BEAST fits.

One potential explanation for this abnormally high extinction is the lack of binary models in the BEAST (adding binaries is in progress). The binary fraction of stars (and subsequent mass ratios) strongly depends on mass and metallicity and is still an active field of research. 
For most binaries, if the mass of the secondary is considerably less than the primary (assuming similar ages and metallicities), the difference between the combined flux versus the primary flux alone will be negligible since $L\propto M^3$. For example, a mass ratio of 10\% will roughly translate to a luminosity ratio of 0.1\%. However, for binaries with larger mass ratios (e.g.~mass ratio $\sim50\%$), the combined increase in magnitude and color can be prominent on a CMD. 

To illustrate this effect, in Figure \ref{fig:binaries}, we overlay PARSEC isochrones (log(Age)$=9.8$ yrs, $Z=20\%\ Z_{\odot}$, distance of 60 kpc) with $A_V = 0$ mag (blue), $A_V = 1$ mag (red), and $A_V = 0$ mag but assuming equal-mass binary companions (mass ratio $= 1$; orange). Both the extinct isochrone (red) and the binary isochrone (orange) overlap the high-extinction sources in the lower main sequence. However, along the RGB, the effects of binaries are degenerate with single-star isochrones. We note that this overlap will vary depending on the binary mass ratio and the filter combinations.

It is possible that the high-extinction sources in the lower main sequences of the SMC are actually the consequence of misclassified binary stars. Future BEAST modeling of binary stars will amend these misclassifications, however, due to the scarcity of these stars, they have relatively little impact on the average extinction of a field; the median extinction per field is on average $0.050 \pm 0.035$ mag and $0.058 \pm 0.053$ mag less dusty in the SMC and LMC, respectively, if we exclude these dimmer sources (F814W $> 22$ mag). As such, we do not believe these sources will have a significant impact on the overall BEAST statistics.

\subsection{Star-Formation History}\label{sec:sfh}

The MCs have a long interaction history which is thought to have generated bursts of star formation when close encounters occurred \citep{weisz2013, patel2020, cohen2024a, cohen2024b}. Ongoing or recent star formation ($\leq 100$ Myrs) can be traced via observables such as H$\alpha$ or UV radiation from massive stars \citep{weisz2012, mcquninn2015, cignoni2019}. However, tracing the star formation history (SFH) of galaxies requires studying all stellar populations. 

Stellar modeling shows how stars with different initial masses and metallicities evolve and change color/magnitude over the course of their lives. On a statistical level, these stellar models allow us to figure out what history of star formation would have produced the distribution of stars observed in a CMD in the present day. Critical to this analysis is the assumption of an initial mass function (IMF), the rate of enrichment (i.e. age-metallicity relation), and the stellar evolution models employed.

The star-formation histories (SFHs) of the SMC and LMC have been extensively studied. Most recently, \citep{cohen2024a, cohen2024b} used two-band optical imaging from the Scylla survey \citep{murray2024b} and archival HST imaging to constrain the SFHs across different regions in the MCs. They calculated the SFHs for each field by modeling the observed CMDs using the software code \texttt{MATCH} \citep{dolphin2002}. One of the main metrics measured from these studies is $\tau_{X}$, the amount of time it takes to form X\% of the total stellar mass (e.g.~$\tau_{50}$ is the lookback time by which 50\% of the cumulative stellar mass of a galaxy has been formed). These metrics are used to make it easier to compare star-formation histories between galaxies of different masses.

The BEAST is not designed to infer the SFH of galaxies, as it is only able to observe the age distribution of stars in the present day (since older high-mass stars will be missing). However, we can still compare the observed age distributions with SFH measurements to validate BEAST age estimates.

In Figure \ref{fig:sfh}, we plot the cumulative distributions of the BEAST stellar age for each Scylla field in the SMC (left; red) and LMC (right; blue), and highlight the cumulative distribution for all sources combined. Markers note the 50th (\texttt{o}), 75th (\texttt{x}), and 90th (\texttt{$\triangle$}) percentiles of the cumulative stellar age distributions based on all HRC sources compared to the median $\pm$ $1 \sigma$ $\tau_{50}$, $\tau_{75}$, and $\tau_{90}$ measurements for each Scylla and METAL field in the SMC and LMC \citep{cohen2024a, cohen2024b}. We also plot the total cumulative star formation history for each galaxy as a grey line (Burhenne et al. 2024, in prep.) and show the cumulative distribution for a constant star-formation rate as a dashed line. 

\begin{figure*}
    \centering
    \includegraphics[width=0.9\textwidth]{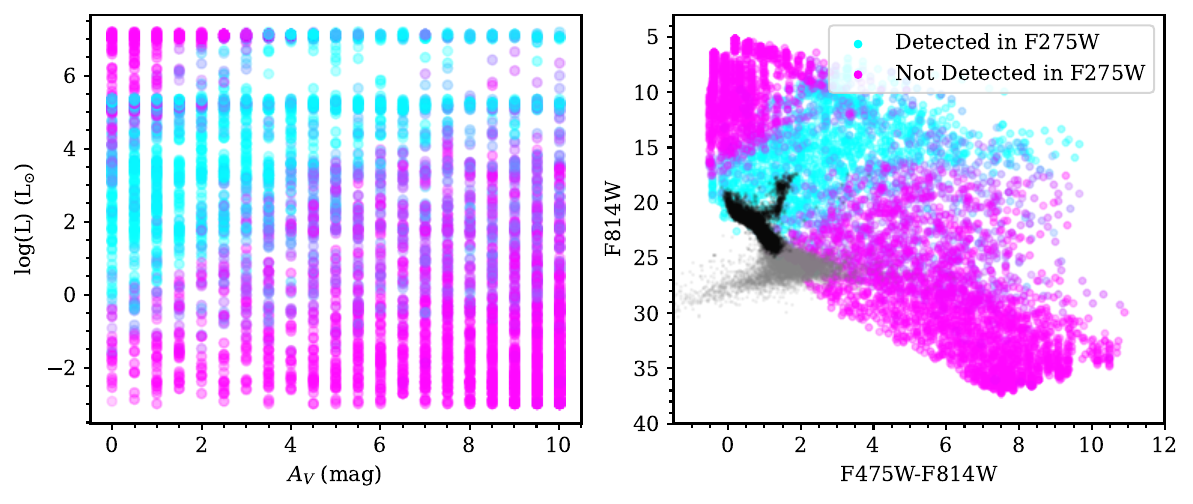}
    \caption{\textbf{Extinction detection limit}: Left: 24,039 artificial star tests (ASTs) from a 6-filter Scylla field in the SMC (15891{\textunderscore}SMC-1588ne-12105) plotted as a function of model extinction and extinguished luminosity. Sources are colored by whether they were detected in the bluest filter F275W (cyan) or not (magenta). With the given exposure times, most artificial stars are undetectable, however, some sources are too bright and saturate. Right: Optical CMD of ASTs overlaid with the actual observed sources in the example Scylla field, with all sources in grey, and high-quality sources in black. Sources in the upper main sequence and RGB are easily observable with 3 mag of extinction. }
    \label{fig:av_detect}
\end{figure*}

The BEAST stellar age distributions show the majority of observed stars are younger than 6-7 Gyrs, with 77.5\% (93.9\%) of stars in the SMC and 57.1\% (78.8\%) of stars in the LMC being younger than 6.0 (7.0) Gyrs, respectively. These findings are in strong agreement with existing measurements of the SFR, which found a burst of star formation around 5-6 Gyrs ago. In the SMC, the median observed stellar age distribution is in strong agreement with the cumulative mass fraction for the first 6 Gyrs. However, we observe a divergence beyond this age limit that is likely caused by either: (1) observational bias of late-stage intermediate-mass stars (i.e. white dwarf stars are likely too dim to be observable); or (2) observational bias of low-mass stars beyond the completeness limits. In the LMC, there is less agreement between the observed stellar age distribution and the cumulative mass fraction, however, the median 50th, 75th, and 90th age percentiles are still within 1-2$\sigma$ of the corresponding $\tau$ measurements.

Comparing $\tau_{50}$ and $A_{50}$ comes with caveats. Importantly, BEAST age distributions are calculated using observed present-day stellar populations, meaning the stellar age distributions will be missing stars that have already died or stars that are too dim/bright to be observable. By comparison, SFH fits take observational incompleteness into account. Additionally, SFH fits usually include some consideration for binary stars, whereas the BEAST assumes a binary fraction of 0 since it only uses single-star models. Intrinsically, SFH fits like $\tau_{50}$ are a measurement of the time it took to form half of the cumulative stellar mass ever formed, whereas $A_{50}$ only measures the median age of the stars observable in the present day. 
In general, $A_{50}$ should be systematically younger than $\tau_{50}$, especially for further lookback times. With all of this in mind, it is rather impressive that the stellar age distributions from the BEAST are in such strong agreement with robust SFH measurements, especially in the SMC. These results ultimately indicate that, for stellar populations, the BEAST fits are capable of accurately recovering the ages of stars.

\subsection{Dust Properties}\label{sec:extinction}

Measuring dust extinction through resolved stellar SED fitting gives us valuable new insight into the gas content of nearby galaxies. In this section, we examine the BEAST $A_V$ measurement more closely and discuss upcoming mapping techniques, define detection limits for high extinction, quantify the average extinction distribution for each galaxy, and compare dust measurements with other ISM tracers available. 

\subsubsection{Mapping Dust Extinction}

Mapping the distribution and composition of dust structures in galaxies is critical for tracing the multi-phase structure of the interstellar medium (ISM). Most dust maps are either constructed using IR dust emission, which relies on assuming a dust emissivity coefficient, or use stellar types like RGB stars to measure a reddening vector in a stellar population \citep{dalcanton2015, ymj2017}, although this method limits the number of viable stellar probes. By comparison, BEAST extinction measurements provide high-resolution probes of the ISM using \textit{all} the stars in a field, not just certain stellar types. In upcoming Scylla publications, we will harness these discrete probes of extinction to infer the underlying distribution of dust within individual fields (Lindberg et al. 2025, in prep). For the purposes of this paper, the main goal is to: (1) quantify how much extinction we can probe with the observed Scylla sources; and (2) compare extinction with other tracers of the ISM.

\subsubsection{Extinction Detection Limit}\label{sec:av_limit}

The maximum extinction we can probe is dependent on both the unextinguished luminosity of stars and the wavelengths observed. Brighter stars can probe larger magnitudes of extinction before the sources fall below observational detection limits. Meanwhile, the effects of extinction are wavelength-dependent and are more prominent in the UV, affecting the F225W, F275W, and F336W bands the most. To quantify the $A_V$ detection limit in the Scylla fields, we use artificial star tests (ASTs) to test how stellar recovery varies as a function of simulated extinction and intrinsic stellar luminosity. 

In Figure \ref{fig:av_detect} (left), we plot the model extinctions and intrinsic luminosities for 24,039 ASTs simulated for 7-filter Scylla field 15891{\textunderscore}SMC-1588ne-12105. Sources were generated using a grid of extinctions ranging from $0.0\mbox{--}10.0$ mag with a grid spacing of $0.5$ mag, whereas luminosities are calculated from the initial stellar mass (which is sampled from a Kroupa IMF) and log-ages ranging from $6.0\mbox{--}10.0$ years with a grid spacing of $1.0$ dex. ASTs were processed in the same pipeline as described in Section \ref{sec:noisemodel}. For each AST, we note whether the star was observed in F275W (cyan) or not (magenta). We find that sources dimmer than $1\ L_{\odot}$ are generally not detected under any observing conditions, a finding corroborated by the lower-mass limit of $\sim0.7\ M_{\odot}$ found in Section \ref{sec:mass}. Conversely, stars with $log(L) > 4\ L_{\odot}$ are too bright and saturated with long Scylla exposure times. However, some of these bright stars become observable as larger amounts of extinction are applied. 

ASTs are generated to span a wide range of observed fluxes and are not inherently representative of the majority of sources observed in the Scylla fields. To understand which stellar populations are observable with larger extinctions, we compare the photometric overlap between the observed sources in the Scylla field and the detected ASTs. In Figure \ref{fig:av_detect} (right), we plot ASTs by their extinguished model photometric in an optical CMD, maintaining the coloring showing where ASTs were detected in F275W (cyan) and not (magenta). We overlay all photometric sources originally observed in the Scylla field (grey), as well as the sources that pass the high-reliability (HRC) quality cuts (black). 

In summary, most luminous stars in the upper main sequence and RGB ($log(L) \approx 2\ L_{\odot}$) are detectable with 3 mag of extinction, well below current extinction estimates for the MCs \citep[$A_V$ = 0.46 and 0.55 mag in the SMC and LMC, respectively;][]{zaritsky2002, zaritsky2004}. Massive stars ($log(L) = 3\mbox{--}4\ L_{\odot}$) provide the greatest opportunity to probe extreme levels of extinction ($A_V = +3\mbox{--}8$ mag). Low detection limits for sources dimmer than $1\ L_{\odot}$ once again evidence that high-extinction low-mass sources discussed in Section \ref{sec:binaries} are likely just misclassified binary stars given that these sources should, in theory, not be detectable with $A_V = 2$ mag.

\begin{figure}
    \centering
    \includegraphics[width=\linewidth]{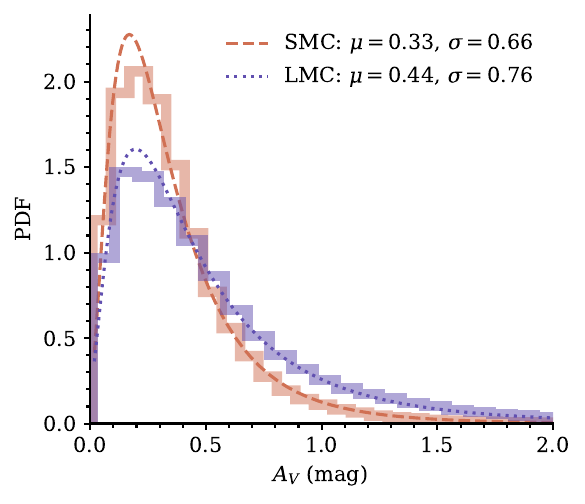}
    \caption{\textbf{Log-Normal Extinction Distribution}: 1D histogram of median BEAST extinctions ($A_V$) for all high-reliability sources in the SMC (red) and LMC (blue) in fields observed with at least 4 filters. Dashed (SMC) and dotted (LMC) lines show the best-fit log-normal functions to each distribution.}
    \label{fig:av_lognorm}
\end{figure}

\subsubsection{Log-Normal Extinction Distribution}\label{sec:av_lognorm}

The ISM is a multi-scale structure with volume-filling atomic gas that cools and coalesces to dense molecular structures that eventually form stars. As a result, the distribution of gas and dust densities in a given area of a galaxy is often well described by a log-normal distribution \citep{vazquez-semadeni1994, ostriker2001, dalcanton2015, Lombardi2015, Schneider2015}, where turbulence is being injected into the medium and only a small fraction of the ISM exists at higher densities. Although we cannot measure the continuous density distributions of the ISM via extinction directly, we can still probe this structure with individual stellar sightlines.

We observe log-normal $A_V$ distributions in both the SMC and LMC, as initially illustrated in Figure \ref{fig:beast}. This is despite the fact that BEAST priors for $A_V$ were flat and covered a wide range of magnitudes ($0\mbox{--}10$ mag). In Figure \ref{fig:av_lognorm}, we characterize these distributions by calculating the best-fit log-normal functions for each galaxy. The SMC is best characterized by a log-normal distribution with $\mu=0.33$ and $\sigma=0.66$ (median $A_V = 0.29$ mag), whereas the LMC is dustier with $\mu=0.44$ and $\sigma=0.76$ (median $A_V = 0.40$ mag). These findings are similar to previous extinction studies in the MCs, which found log-normal distributions with median values of $0.46$ mag in the SMC \citep[see Figure 19 in][]{zaritsky2002} and $0.55$ mag in the LMC \citep[see Figure 9 in][]{zaritsky2004}.

\begin{figure}
    \centering
    \includegraphics[width=0.94\linewidth]{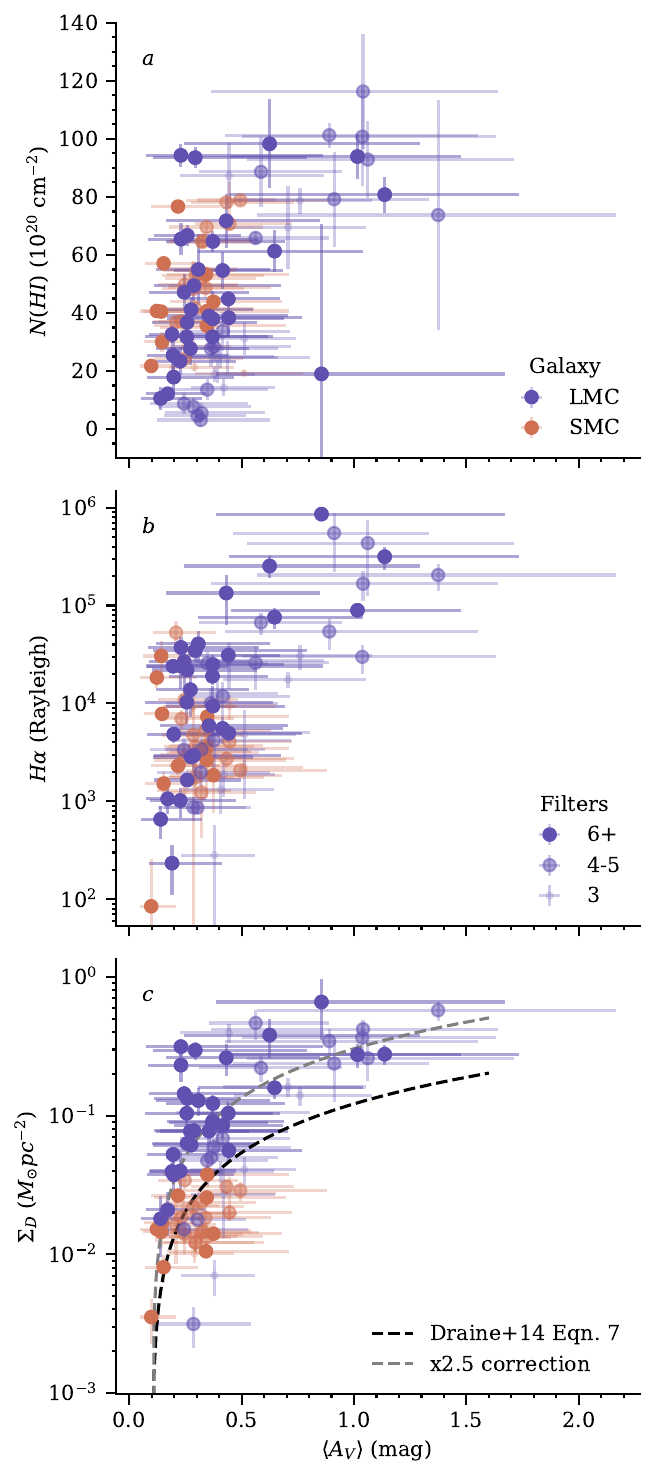}
    \caption{\textbf{Extinction vs.~ISM tracers:} Median BEAST $A_V$ compared to other ISM tracers per Scylla field: (a) atomic hydrogen \citep[N($HI$)][]{kim2003, pingel2022}, (b) ionized hydrogen \citep[$H\alpha$,][]{Gaustad2001}, and (c) FIR dust emission \citep[$\Sigma_D$,][]{clark2021}. Uncertainties denote the standard deviation of extinction or tracer values found within each field. Panel c shows the theoretical dust surface density as a function of $A_V$ based on Eqn.~7 in \citet[][black dashed line]{draine2014}, and the same equation but with a 2.5 correction factor as noted in \citet[][grey dashed line]{dalcanton2015}. A 0.1 mag offset is included along the x-axis to account for foreground extinction.}
    \label{fig:av}
\end{figure}

\subsubsection{Comparison with ISM Tracers}\label{sec:ism_tracers}

The various phases and composition of gas say a lot about the environments present in galaxies. 
Atomic hydrogen gas ($HI$) is the most prevalent phase of gas in the ISM, and is especially dominant in dwarf galaxies like the SMC \citep{leroy2007, jameson2016}. We observe neutral atomic hydrogen via 21-cm emission which allows us to measure the abundance of atomic hydrogen and its radial velocity. As gas cools and condenses, shielding from ionizing radiation allows for the formation of molecules that can be used to trace denser regions prerequisite to star formation. While molecular hydrogen, $H_2$, is the most prevalent molecule, its weak emission means it is often underdetected. As a compromise, carbon monoxide (CO) is often used as an alternative molecular gas tracer since its dipole moment produces rotational transitions (e.g.~J=1-0 at $\lambda$=2.6 mm) observable at low temperatures ($\sim$5 K), although its usage requires significant assumptions \citep[e.g.,][]{madden1997, grenier2005, wolfire2010, jameson2018, madden2020}. Within star-forming regions, UV radiation from young stars photoionizes nearby hydrogen, which recombines and de-excites, producing $H\alpha$ radiation in the process ($\lambda = 656.28$ nm).

Dust is well mixed with atomic and molecular gas \citep{Boulanger1996, Rachford2009}, meaning dust can be used to trace the combined atomic and molecular gas column densities along lines-of-sight. Dust can either be observed through extinction, as we do in this paper via the BEAST, or through emission in the far infrared (FIR) as dust particles are heated by the interstellar radiation field/nearby stars and re-emit energy in the FIR/sub-mm regime \citep{gordon2014}. 

In the following section, we look at how three separate ISM tracers correlate with dust extinction across the fields: atomic gas via $HI$ emission, ionized gas via H$\alpha$ emission, and dust via FIR emission. We chose to not make comparisons with molecular gas tracers since, to date, there are no surveys that observe the entire LMC or SMC. 
These tracers were all observed at different resolutions, so we opt to convolve all emission measurements to the resolution of a single \textit{HST} Scylla field: $\sim160^{\prime\prime}$ for fields with only optical and UV coverage with WFC3 UVIS, or $\sim130^{\prime\prime}$ for fields with IR coverage from WFC3 IR; and only compare with BEAST extinction values averaged across each field. 
In Figure \ref{fig:av}, we show how each of these tracers correlates with the median BEAST-derived extinction measurements.

\paragraph{Atomic Hydrogen}

To trace atomic hydrogen ($HI$) across the SMC, we use 21-cm emission from the recent GASKAP-HI collaboration \citep{pingel2022}. These $HI$ maps were constructed using new observations from the ASKAP telescope \citep{hotan2021} combined with single-dish observations from the GASS survey \citep{mccluregriffiths2009, kalberla2010, kalberla2015}, producing $HI$ emission data cubes with an angular resolution of 30$^{\prime\prime}$, or $\sim$10 pc assuming a distance of 50-60 kpc. For the LMC, we use combined $HI$ emission from the Australia Telescope Compact Array (ATCA) and Parkes receiver \citep{kim2003} with an angular resolution of 60$^{\prime\prime}$. We calculate the median $HI$ emission across each Scylla field. 

In Figure \ref{fig:av}a, we find a linear correlation between $HI$ and extinction, up until $A_V > 0.8$ mag, after which $HI$ plateaus in LMC fields. These fields with higher average extinction are all located within the 30 Doradus star-forming region of the LMC and are likely dominated by molecular gas. Other extinction surveys find similar trends when comparing $A_V$ and $HI$-based $A_V$ measurements at $40^{\prime\prime}$ resolution in the LMC \citep[see Figure 12 in][]{choi2018}.

In the Milky Way, the transition from $HI$-to-$H_2$ observed to occur at around $A_V = 0.3$ mag \citep{liszt2014b}, meaning, for fields below this transition threshold, the relationship between $HI$ column density and $A_V$ should become linear. Beyond this threshold, the relationship should flatten as $A_V$ continues to trace the total column density of neutral gas with contributions from $H_2$, while $HI$ remains constant. The transition threshold from $HI$-to-$H_2$ should be independent of metallicity, so this inconsistency between the Milky Way and the LMC is likely due to under-resolving the molecular regions within Scylla fields in this analysis. Further work at sub-field resolutions is needed to resolve these inconsistencies.

\paragraph{Ionized Hydrogen}

H$\alpha$ broadly traces the amount of ionizing radiation from young stars, and should therefore trace regions of recent star formation. 
To trace warm ionized interstellar gas, we use $H\alpha$ emission from the Southern H-Alpha Sky Survey Atlas \citep[SHASSA,][]{Gaustad2001}, which observed the southern hemisphere sky using the Cerro Tololo Inter-American Observatory (CTIO) at 48$^{\prime\prime}$ pixel-resolution, or $\sim$13 pc assuming a distance of 50-60 kpc. Similar to $HI$, we calculate the median $H\alpha$ emission for each Scylla field.

Figure \ref{fig:av}b shows a logarithmic correlation between $A_V$ and H$\alpha$ emission for fields in the LMC, whereas in the SMC, there is very little correlation. 
Fields with higher levels of extinction tend to be correlated with higher levels of $H\alpha$ emission in the LMC. This is not surprising given that these high-extinction fields are all located in 30 Doradus, the most active star-forming region in the Local Group \citep{sabbi2013}. Perhaps more surprising is the fact that fields with $A_V < 0.5$ mag span several orders of magnitude in $H\alpha$ emission, an indication that either: (1) dustier star-forming structures constitute a relatively small area within any given field and are therefore not contributing significantly to the median extinction; (2) any dusty surrounding material has been destroyed by the ionizing radiation \citep[e.g.,][]{kruijssen2014}. Once again, research on sub-field resolutions will be necessary to constrain the two scenarios.

\paragraph{FIR Dust Emission}

To compare dust extinction with FIR dust emission-derived dust surface densities, we use a new combination of FIR emission observations of the MCs \citep{clark2021}. These maps were constructed using re-reduced high-resolution Herschel data from the Herschel Inventory of the Agents of Galaxy Evolution \citep[HERITAGE;][]{meixner2013} combined with lower-resolution data from all-sky surveys like Planck \citep{planck2011}, the Infrared Astronomical Satellite \citep[IRAS][]{neugebauer1984}, and the Cosmic Background Explorer \citep[COBE][]{boggess1992, silverberg1993} which capture extended dust emission. 

In Figure \ref{fig:av}c, we plot dust surface density ($\Sigma_D$) as a function of the median $A_V$ for each Scylla field. We compare this correlation with previous estimates of FIR-derived dust surface densities as described by Eqn. 7 in \citet[][]{draine2014}: 

$$A_V = 0.74 \left( \frac{\Sigma_D}{10^{-1} M_{\odot} \mathrm{pc}^{-2}} \right) \mathrm{mag}$$

We calculate the theoretical dust surface density based on the range of observed BEAST $A_V$ and overlay the results as a dashed black line, including a 0.1 mag offset along the x-axis to account for potential foreground extinction. Although the relation between $A_V$ and $\Sigma_D$ is similar, there is a notable offset. This is not the first instance of discrepancy found: \citet{dalcanton2015} found a factor of $\sim2.5$ difference between their extinction inferred from CMD fitting versus dust surface densities from FIR dust emission modeling in M31. Similarly, \citep{planck2016} found a factor of $\sim2$ factor difference between \citet{draine2007} extinction estimates versus quasar $A_V$ measurements, and \citet{ymj2021} found a factor of 1.8 difference between CMD extinction measurements in the SMC versus dust surface densities derived measured from FIR emission modeling. When we apply a correction factor of 2.5 to $\Sigma_D$ (grey dashed line), the model is now within $1\sigma$ of the observed emission, indicating that BEAST $A_V$ estimates are in fact in agreement with existing FIR-derived dust surface density measurements.

For all ancillary ISM tracers, we find positive correlations with the median BEAST-derived extinction measurements in the MCs. However, in the SMC, these correlations are much less pronounced owing to the smaller amounts of dust in the SMC and resolution effects. For both galaxies, these correlations indicate that BEAST-derived extinction measurements are broadly capturing the average ISM content in each Scylla field. Future Scylla publications will expand this investigation by comparing the BEAST-derived extinction parameters ($A_V$, $R_V$, and the presence of the 2175~\r{A} bump) at the original resolutions of the ancillary ISM tracers, and use extinction to quantify the fraction of CO-dark molecular gas as a function of metallicity between the LMC and SMC (Murray al. 2025, in prep). On their own, extinction measurements can be used to construct high spatial resolution extinction maps (Lindberg et al. 2025, in prep).

\section{Summary}
\label{sec:summary}

In this work, we present stellar and line-of-sight dust fits to over 1.5 million sources in the MCs. From these initial fits, we define quality cuts to extract 550,000 high-reliability sources in the SMC and LMC (257,037 and 309,359 sources, respectively). Sources were observed in multiple \textit{HST} filters spanning from the UV to the IR as part of the Scylla survey \citep{murray2024b}. 
Our data consists of 33 fields in the SMC and 56 fields in the LMC observed with \textit{HST WFC3} as part of the Scylla \citep{murray2024b} and METAL \citep{romanduval2019} surveys. Scylla fields were observed with filters in the following priority order when scheduling allowed: F475W, F814W, F336W, F275W, F110W, F160W, F225W. We only fit fields with at least 3-filter coverage. 

We use the BEAST \citep[][]{gordon2016} to characterize the multi-band SEDs (3-7 filter coverage), providing us with probabilistic estimates of initial mass, age, metallicity, distance, extinction ($A_V$), total-to-selective extinction ($R_V$), and extinction curve (Milky Way-like or SMC-like without a 2175~\AA) for each individual star. We compare these measurements to existing literature on the stellar content of the MCs, their star-formation history, and the ISM. Our main findings are as follows:

\begin{enumerate}
    \item \textit{Stellar Mass} - We observe stars with initial masses as low as $0.6\ M_{\odot}$, and find that initial mass estimates are in agreement with PARSEC stellar evolutionary tracks \citep{bressan2012}. (\S\ref{sec:mass}; Figures \ref{fig:beast_cmd_lmc}, \ref{fig:beast_cmd_smc} and \ref{fig:iso}) 
    \item \textit{Binary Stars} - Binary stars are a potential source of high-extinction misclassifications, especially for low-mass sources along the main sequence where binary models populate the same CMD space as 1 mag extinction. However, even if these potential binary sources were misclassified as highly extinguished, they would only increase the average extinction in each galaxy by $\sim0.05$ mag. (\S\ref{sec:binaries}; Figure \ref{fig:binaries})
    \item \textit{Star Formation History} - BEAST stellar age distributions are in strong agreement with star formation history measurements for the past 6 Gyrs \citep{cohen2024a, cohen2024b}. The majority of observed stars are fit to be younger than 6-7 Gyrs, with 77.5\% (93.9\%) of stars in the SMC and 57.1\% (78.8\%) of stars in the LMC being younger than 6.0 (7.0) Gyrs, respectively. (\S \ref{sec:sfh}; Figure \ref{fig:sfh})
    \item \textit{Extinction Limits} - Most upper main sequence and RGB/RC stars are detectable with up to 3-mag of extinction. Sub-solar mass stars are not detectable with more than $\sim0.5$ mag of extinction. (\S \ref{sec:av_limit}; Figure \ref{fig:av_detect})

    \item \textit{Extinction Distribution} - Extinction distributions for all sources in both the SMC and LMC follow log-normal functions, with $\mu=0.33$ and $\sigma=0.66$ (median $A_V = 0.29$ mag) in the SMC, and $\mu=0.44$ and $\sigma=0.76$ (median $A_V = 0.40$ mag) in the LMC. (\S \ref{sec:av_lognorm}; Figure \ref{fig:av_lognorm})
    
    \item \textit{Gas Tracers} - Field-averaged BEAST $A_V$ measurements are positively correlated with other ISM tracers like atomic hydrogen ($HI$) and ionized hydrogen ($H\alpha$). $A_V$ is linearly correlated with $HI$ until $\sim0.8$ mag. Above this limit, observed fields are mainly located in the 30 Doradus region of the LMC, meaning the gas content is likely predominantly in the molecular phase. $A_V$ is logarithmically correlated with $H\alpha$ emission, but low-$A_V$ fields ($A_V < 0.5$ mag) span several orders of magnitude in $H\alpha$ emission ($10^2\mbox{--}10^5$ Rayleigh). (\S \ref{sec:ism_tracers}; Figure \ref{fig:av})
    
    \item \textit{FIR-based Dust Surface Densities} - We convert observed BEAST $A_V$ measurements to predicted dust surface densities using the \citet{draine2014} model and find that the $A_V$-based dust surface densities are a factor of $\sim$2.5 lower than observed FIR-based dust surface densities. This correction factor is similar to other studies, both in the SMC \citep[e.g.~][]{ymj2021}, and other galaxies like M31 \citep[e.g.~][]{dalcanton2015}. (\S \ref{sec:ism_tracers}; Figure \ref{fig:av})

    \item \textit{Future Work} - The results presented in this paper are predominantly focused on galaxy-/field-scale statistics of BEAST fits. In upcoming Scylla ISM publications we plan to focus on sub-field scales by: (1) constructing parsec-scale extinction maps; (2) investigating the source of sub-field variations in $R_V$; and (3) constraining the fraction of CO-dark molecular gas in the SMC and LMC.  
\end{enumerate}

\facilities{HST (WFC3/IR), HST (WFC3/UVIS), XSEDE \citep{xsede2014}}

\software{DOLPHOT \citep{dolphin2002}, BEAST \citep{gordon2016}, astropy \citep{2013A&A...558A..33A, astropy}, numpy \citep{harris2020array}, scipy \citep{Virtanen2020}, matplotlib \citep{Hunter2007}}

\acknowledgments{This research is based on observations made with the NASA/ESA Hubble Space Telescope obtained from the Space Telescope Science Institute, which is operated by the Association of Universities for Research in Astronomy, Inc., under NASA contract NAS 5–26555. These observations are associated with programs 15891, 16235, and 16786. This research has made use of NASA’s Astrophysics Data System. All of the data presented in this paper were obtained from MAST at the Space Telescope Science Institute. The specific observations analyzed can be accessed via \dataset[https://doi.org/10.17909/8ads-wn75]{https://doi.org/10.17909/8ads-wn75}. Support to MAST for these data is provided by the NASA Office of Space Science via grant NAG5–7584 and by other grants and contracts.

This project made use of data from the Southern H-Alpha Sky Survey Atlas (SHASSA), which is supported by the National Science Foundation

This work used Bridges-2 at Pittsburgh Supercomputing Center (PSC) through allocation \#AST190023P from the Advanced Cyberinfrastructure Coordination Ecosystem: Services \& Support (ACCESS) program, which is supported by National Science Foundation grants \#2138259, \#2138286, \#2138307, \#2137603, and \#2138296. This work used the Extreme Science and Engineering Discovery Environment (XSEDE), which is supported by National Science Foundation grant number ACI-1548562. Specifically, it used the Bridges-2 system, which is supported by NSF award number ACI-1928147, at the PSC.
}

\bibliographystyle{aasjournal}
\bibliography{main}

\appendix

\section{UV and IR BEAST CMDs \label{apx:cmd}}

\begin{figure*}[h]
    \centering
    \includegraphics[width=\textwidth]{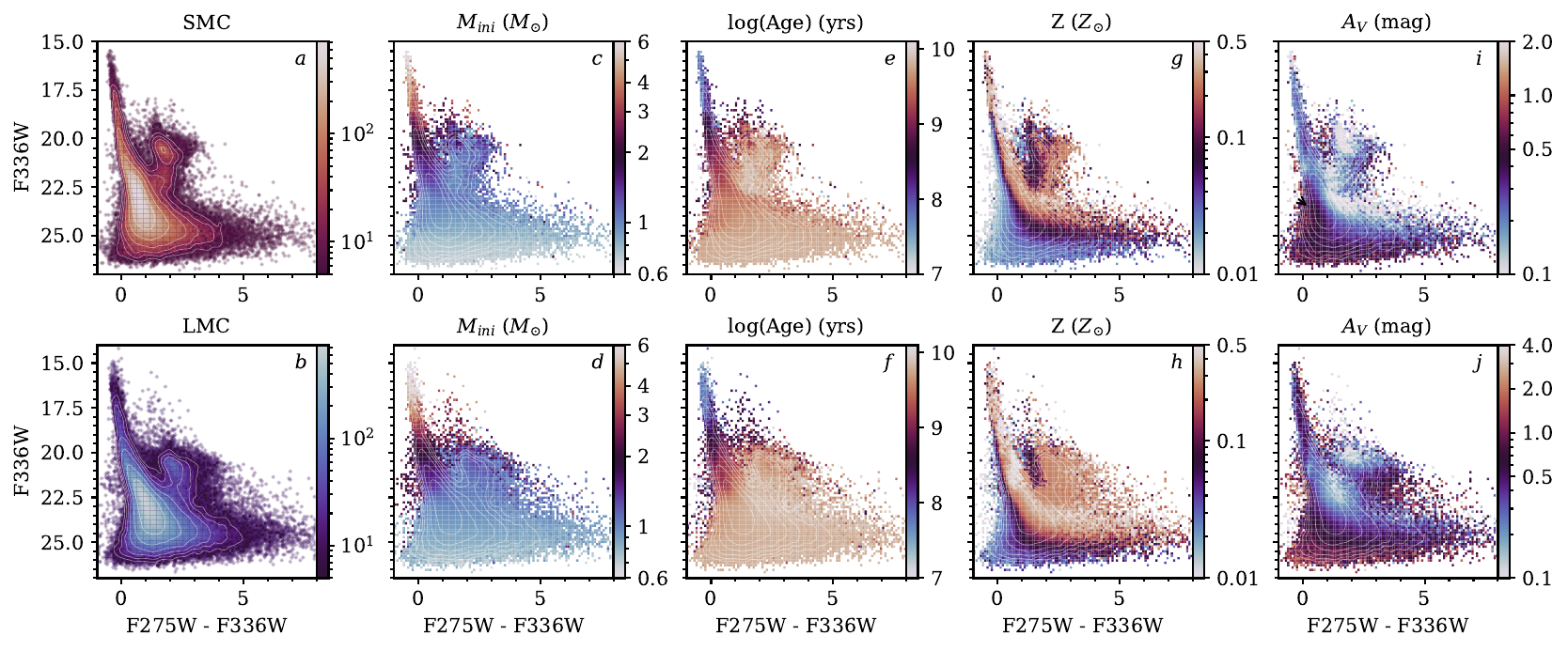}
    \includegraphics[width=\textwidth]{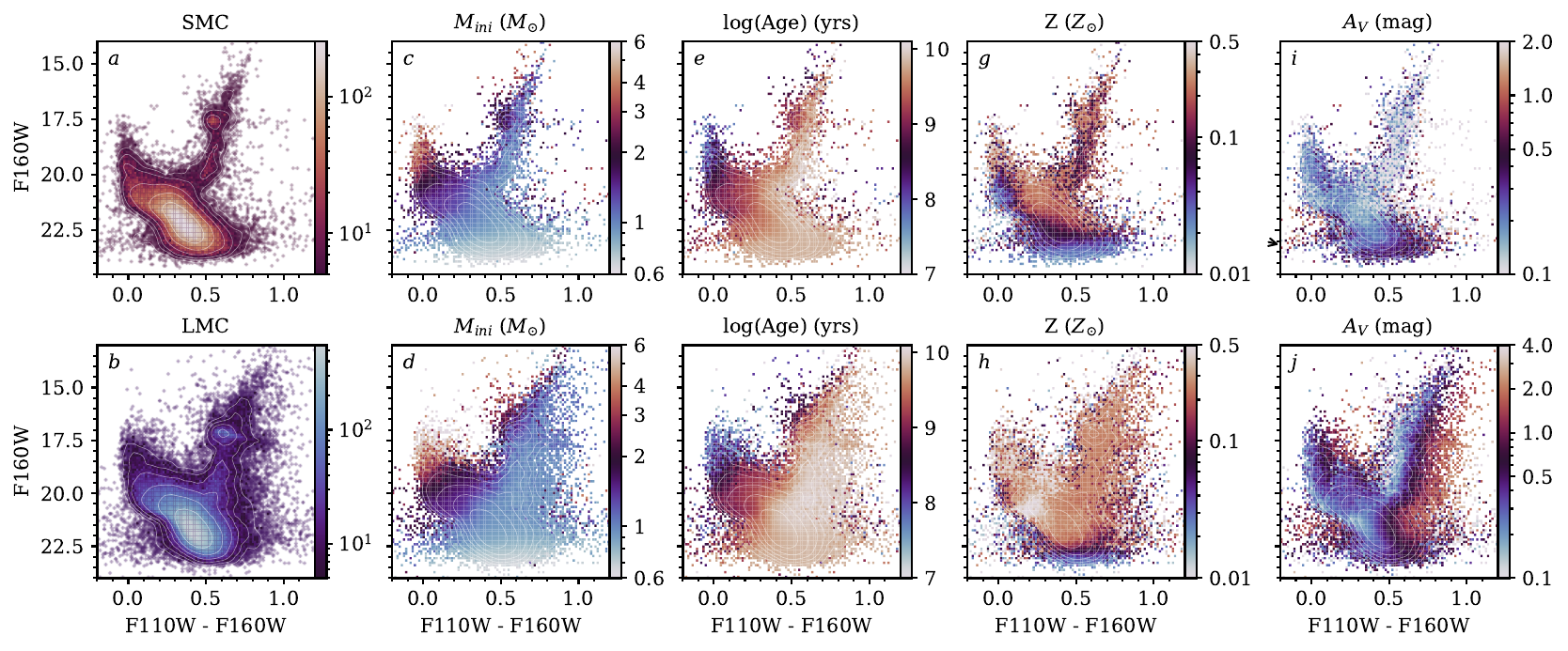}
    \caption{\textbf{UV and IR versions of Figures \ref{fig:beast_cmd_lmc} and \ref{fig:beast_cmd_smc}}: UV (F275W and F336W) and IR CMDs (F110W and F160W) for \textit{all} high-reliability sources in the SMC (top; $n_{\mathrm{UV}}=139,551$ and $n_{\mathrm{IR}}=59,231$) and LMC (bottom; $n_{\mathrm{UV}}=188,555$ and $n_{\mathrm{IR}}=137,023$). Left columns ($a, b$): Density distribution of sources as colored 2D histograms, and outlined in logarithmic levels using kernel-density estimates (KDE). KDE density outlines (white) are propagated to all other columns. All other columns ($c$-$j$): Spatially binned ($100\times100$) medians for notable BEAST parameters: initial mass ($M_{ini}$), age, metallicity ($Z$), and extinction ($A_V$). }
    \label{fig:beast_cmd_uv}
\end{figure*}

In Figure \ref{fig:beast_cmd_uv}, we show the UV and IR versions of the optical CMDs in Figures \ref{fig:beast_cmd_lmc} and \ref{fig:beast_cmd_smc}.


\end{document}